\documentstyle[preprint,aps]{revtex}

\draft
\tightenlines

\input{psfig}

\begin{document}

\title{Irreducible~pionic~effects~in~nucleon-deuteron~scattering~below~20~MeV}

\author{ L. Canton$^{1}$, W. Schadow$^{2,}$\footnote {Permanent
address: Cap Gemini Ernst \& Young, Hamborner Str. 55, 40472
D\"usseldorf, Germany}, and J. Haidenbauer$^{3}$ }
\address{ $^1$Istituto Nazionale di Fisica
Nucleare, Sez. di Padova, Via F. Marzolo 8, Padova I-35131, Italy \\
$^2$ Dipartimento di Fisica dell'Universit\`a di Padova, Via F.
Marzolo 8, Padova I-35131, Italy \\
$^{3}$Forschungszentrum J\"ulich, IKP, D-52425 J\"ulich, Germany}

\date{12 November, 2001}

\maketitle

\begin{abstract}
The consequences of a recently introduced irreducible pionic effect in
low energy nucleon-deuteron scattering are analyzed.  Differential
cross-sections, nucleon (vector) and deuteron (vector and tensor)
analyzing powers, and four different polarization transfer
coefficients have been considered.  This 3$N$$F$-like effect is
generated by the pion-exchange diagram in presence of a two-nucleon
correlation and is partially cancelled by meson-retardation
contributions. Indications are provided that such type of effects are
capable to selectively increase the vector (nucleon and deuteron)
analyzing powers, while in the considered energy range they are almost
negligible on the differential cross sections.  These indications,
observed with different realistic nucleon-nucleon interactions,
provide additional evidences that such 3$N$$F$-like effects have indeed
the potential to solve the puzzle of the vector analyzing powers.
Smaller but non negligible effects are observed for the other spin
observables.  In some cases, we find that the modifications introduced
by such pionic effects on these spin observables (other than the
vector analyzing powers) are significant and interesting and could be
observed by experiments.
\end{abstract}

\vspace{2mm}

\section{Introduction}

Three-nucleon forces as truly genuine interactions are not made by
nature~\cite{Glockle84,Sauer92}. Rather, they are artifacts of
theoreticians who decided to treat nuclear dynamics within a
restricted Hilbert space.

A sufficiently detailed picture for most low- and
intermediate-energy phenomena in nuclear dynamics should involve
as constituents the nucleons, isobars and mesons (pions and
heavier bosons).  In doing so, all subnucleonic degrees of
freedom, like quarks and gluons, are transformed into the notion
of collective motion of hadronic clusters, which correspond to
baryons and mesons. Such a system does not need the concept of
two-, three-, and many-nucleon forces, since the mechanisms of
meson emission and absorption, the ones that mediate the nuclear
interactions, appear in the theory explicitly.

For systems with more than two nucleons, not all the relevant hadronic
degrees of freedom can be traced back explicitly in practical
calculations, and one is faced with the problem to restrict the
Hilbert space. This restriction generates three- (or many-) nucleon
forces. And, most importantly, three-nucleon forces of different
structure and of different physical interpretation arise depending on
the different theoretical strategies which have been adopted to
restrict the original hadronic system into the simplified, more
tractable, Hilbert space of few-nucleon systems.

A paradigmatic example that clarifies this situation is the one
discussed in Ref.~\cite{Glockle84} where two different strategies for
describing the three-nucleon force are considered.  Both approaches
make the assumption that meson-exchange phenomena occur
instantaneously and hence they both ignore any additional mesonic
aspects brought in by the dynamics of the coupled meson-baryon
system. The difference between the two considered approaches is that
one of the two treats explicitly the non-nucleonic degree of freedom
of the $\Delta$ isobar, thereby allowing the isobar to propagate over
a finite time interval, while the other deals with nucleonic degrees
of freedom only (the standard nuclear theory approach) which implies
that one has to consider effective three-nucleon interactions
explicitly, including those effective contributions for {\em not}
allowing the isobar to propagate freely into the system. Following the
first strategy, one ends up dealing with effective three-nucleon force
(3$N$$F$) diagrams which are energy dependent due to the isobar
propagation, and in addition one finds that, for consistency, also the
two-nucleon interaction is modified in an energy dependent way because
of the the presence of the third nucleon.  The two-nucleon interaction
in particular is modified in those contributions where an intermediate
isobar propagation occurs, since the $\Delta$ is allowed to adjust to
the nuclear medium by the dynamical approach.  This secondary effect
on the 2$N$ force (sometimes called ``dispersive 2$N$ effect'')
counteracts the contribution due to the $\Delta$-mediated
2$\pi$-exchange 3$N$$F$, and the physical interpretation of its origin
is well understood in the approach with explicit $\Delta$ isobars.

In contrast, the second strategy uses two- and three-nucleon
interactions which are not allowed to adjust to the surrounding
nuclear medium, because of the freezing of all the non nucleonic
degrees of freedom. Conceptually, then it is very difficult to explain
the origin of the dispersive 2$N$ counter-terms, and the approach is
amended to revive at least part of such contributions in an averaged
way by introducing effective 3$N$$F$ counterterms of short-range type.

This paradigmatic example suggests that in the standard
``few-nucleon'' approach, with its restricted Hilbert space
consisting of nucleons only, some additional 3$N$$F$-like diagram
could be missing and the most viable way to reveal these new terms
is to work-out the nuclear dynamics starting from an enlarged
space where the system has to be treated dynamically also in
presence of mesons, especially of the pionic degrees of freedom.
If the origin of the 3$N$$F$ effects is due to the restriction to
the nucleonic Hilbert space, then we have to consider carefully
how we theoretically reach this description, in order to pin-down
the 3$N$$F$ contributions. This is one lesson we must learn from
the dynamical treatment of the $\Delta$ isobar in the
three-nucleon system.

The first non-nucleonic ingredient that plays a crucial role for
nuclear systems is the pion, and it is natural to aim at a theoretical
formulation that couples dynamically this meson to the nucleonic
degrees of freedom. Such a formulation would then provide a combined
description of nuclear systems at low- and intermediate
energies. Pions are produced indeed and observed by experiments at
intermediate energies and their dynamical role in these nuclear
systems can hardly be ignored.

With two-nucleon systems, various approaches to include the
effects of a dynamical pion in a non-perturbative fashion have
been developed and analyzed~\cite{TRABAM}, and many theoretical
problems like relativistic aspects, or the proper treatment of the
nucleon renormalization effects, or finally the consistent
treatment of meson-exchange diagrams related by different time
orderings, have been extensively discussed~\cite{Garcilazo90}. Not
all these problems have been solved in a systematic manner since
these $\pi$$NN$ approaches aim to include just one dynamical pion
in the theory, thus ruling out more complex situations with
multipionic intermediate states.  Stated in other words, these
techniques allow to include one single dynamical pion in the 2$N$
system, but then one has to face the conceptual problems arising
because the states with more than one pion at the same time are
ruled out from the dynamical equations.  Still, in spite of these
problems, the generalization of such type of approaches to the
three-nucleon system allows to reveal possible new dynamical
aspects of the pion that cannot be observed by freezing out from
the very beginning {\it all} mesonic degrees of freedom, into
instantaneous meson-exchange potentials.

The dynamical equations for the coupled $NN$-$\pi$$NN$ problem
have been generalized to the 3$N$ system in Ref.~\cite{Canton98}.
The method is based on the rigorous
Yakubovski-Grassberger-Sandhas~\cite{Yakubovski67,Grassberger67}
four-body theory, extended to the $\pi$$NNN$ system with an
absorbable pion.  Previously, this problem has been analyzed with
few-body techniques for quite a few
years~\cite{Avishai83,Cattapan93,Cattapan94-97}.  The solution of
the resulting set of 21$\times$21 coupled equations represents a
formidable task, and for practical reasons an approximation scheme
to reduce the complexity of the original equations to an
approximated but more tractable form has been
developed~\cite{Canton01-a}.  The approximation scheme builds up
the complete dynamical solution starting from the zeroth-order
solution, given by the standard quantum-mechanical
Alt-Grassberger-Sandhas (AGS) equation~\cite{AGS}. The strategy
that emerges from this approximation scheme consists in treating
the normal 2$N$ correlations ``exactly'' via Faddeev-like methods,
while the additional mesonic aspects which cannot be adequately
described by a conventional 2$N$ potential have to be incorporated
directly into the dynamical equation as corrections ({\it i.e.}
through an underlying perturbative-iterative expansion). This is
consistent with the findings of the approaches based on chiral
perturbation theory (ChPT) which predict that 3$N$$F$ effects are
small~\cite{Weinberg92}.

To the lowest order, the approximation scheme ends up with three
different types of irreducible 3$N$$F$ diagrams which have to be
incorporated in the dynamical equations according to the prescriptions
given in Ref.~\cite{Canton01-a}.  Clearly, all three types of diagrams
are related to specific aspects of the pion dynamics which cannot be
incorporated in the description of a purely nucleonic system
interacting through a pair-wise potential.

Amongst the three types of diagrams, one can easily recognize the
extensively studied 2$\pi$-3$N$$F$ diagram. This diagram has led
to various models of 3$N$ potential amongst which we recall two
historical representations, Tucson-Melbourne~\cite{Coon79} and
Brasil~\cite{Robilotta83}.  These potentials use the $\pi$$NN$
vertices and the off-shell extrapolated $\pi$$N$ amplitudes as
inputs. It is important to observe that in the construction of
these potentials certain diagrams have to be subtracted
explicitly, to avoid double countings.  The other two types of
diagrams that emerge at the lowest order involve the intermediate
formation of a 2$N$ correlation while one pion is being exchanged,
and finally the intermediate formation of a 3$N$ correlation,
during the exchange process.  The last type of diagram has been
discussed in the previous literature only
occasionally~\cite{Haberzettl94}, and to our knowledge, its effect
has not yet been estimated quantitatively in realistic situations,
although its relevance might not be so important because the
probability that a full 3$N$ correlation is formed during the
meson-exchange process is expected to be small.

The diagram involving the 2$N$ correlation during a pion-exchange
process has been discussed in Ref.~\cite{Canton00}, and the
consequent appearance of an irreducible 3$N$ operator with
tensor-like structure has also been observed therein (see also
Refs.~\cite{Canton01-b,Canton01-c}).  In the construction of this
operator, certain classes of subdiagrams have to be subtracted,
because of the presence of a cancellation effect which has been
observed in the literature quite a few years
ago~\cite{Yang86,Coon86}. The connection between the nature of
these cancellations and their implications to the ChPT approach
has been discussed in Ref.~\cite{Friar94}.  A similar cancellation
effect has been also observed - in leading order - in effective
nuclear forces based on chiral lagrangians and constructed with
the method of unitary transformation~\cite{Epelbaoum98}.  As has
been discussed in Refs.~\cite{Canton00,Canton01-b,Canton01-c},
however, such subtraction leads to a cancellation effect which is
only partial if the original 3$N$$F$ diagram has been derived
within a dynamical approach which leads to an energy dependent
3$N$$F$-like operator, and where the meson propagates for a
sufficiently extended time to allow the intermediate formation of
a correlated 2$N$ pair. That might correspond to the inclusion of
a series of non-vanishing higher-order terms in the chiral
expansion.  In the standard approach which uses instantaneous two-
and three-nucleon interactions, the freezing out of the mesonic
degrees of freedom does not contemplate the occurrence of such a
3$N$$F$ effect.  This situation presents interesting similarities
with respect to that already encountered with the dispersive 2$N$
counter-terms, which are properly taken into account only when the
$\Delta$ degrees of freedom are treated explicitly in the
dynamical equation.

In Ref.~\cite{Canton01-b}, the consequences of considering this
type of one-pion-exchange 3$N$$F$ diagram (OPE-3NF) in the 3$N$
equation have been studied.  It was shown that this diagram has
the potential to modify considerably the vector analyzing powers
$A_y$ without affecting appreciably the differential cross
section, and could therefore solve the long standing puzzle of the
vector analyzing powers in nucleon-deuteron scattering below 30
MeV. On the other hand, other, alternative explanations have been
suggested in the literature. Effective three-nucleon potentials
constructed with a combination of short-range and pion-range terms
have been considered first under the point of view of the
meson-exchange picture~\cite{Coon95}, and later reconsidered under
the framework of chiral perturbation theory~\cite{Huber01}, since
a non-negligible role for these terms is predicted.  A
phenomenological spin-orbit three-body operator has also been
introduced~\cite{Kievsky99} in order to improve the vector
analyzing powers.  Another explanation~\cite{Sammarruca00}
advocates the Brown-Rho scaling-with-density
hypothesis~\cite{Brown91} in the three-nucleon system.  Because of
this scaling, the 2$N$ potential has been modified specifically in
the triplet P waves by reducing the scalar- and vector-meson
masses to approximately 95\% of their free-space value, thereby
enhancing the spin-orbit term of the density-dependent 2$N$
potential with respect to free space.  This selective dependence
of the $A_y$ puzzle on the triplet P-waves of the 2$N$ force
suggested also that modern phase-shift analysis might have not yet
been settled to the correct parameters for these states in the
low-energy domain~\cite{Tornow98}.  But it has also been
argued~\cite{Huber98} that the required changes of the free-space
2$N$ potential have to be exceedingly drastic for the
one-pion-exchange contribution, which on the contrary is well
established.

In the present paper, we have extended the study of
Ref.~\cite{Canton01-b} about the irreducible pionic effect implied by
the OPE-3NF diagram by considering such an effect with three different
2$N$ potentials.  In all cases we were able to demonstrate that this
irreducible pionic effect has the potential to solve the puzzle for
the low-energy vector analyzing powers with negligible effects for the
differential cross sections and minor effects for the other spin
observables. We have also widened the comparison with experimental
data by including differential cross-sections, vector and tensor
analyzing powers, and spin-transfer coefficients at various energies
below 20 MeV.

The basic structure of the OPE-3NF diagram is recalled in
Sect.\ref{Theory}.  Comparison between theory and experiments is made
in Sect.~\ref{Results}.  Conclusions are derived in
Sect.~\ref{Conclusions}.

\section{Theory}
\label{Theory}

Following Refs.~\cite{Canton98,Canton01-a,Canton00,Canton01-b}, we
have incorporated directly into the Faddeev-AGS three-nucleon equation
the irreducible effects generated by the one-pion-exchange mechanism
in presence of a nucleon-nucleon correlation.  The detailed expression
has been discussed previously in Ref.~\cite{Canton00}. It is reported
here for convenience:
\begin{eqnarray}
\label{OPE3NF}
V^{3N}_3({\bf p,q,p',q'};E) &=&
 {f_{\pi NN}^2(Q)\over m_\pi^2}{1\over (2\pi)^3}
 \\ && \times
  \!\left [
{
({\mbox{\boldmath $\sigma_1$}}\cdot{\bf Q})
({\mbox{\boldmath $\sigma_3$}}\cdot{\bf Q})
({\mbox{\boldmath $\tau_1$}}\cdot {\mbox{\boldmath $\tau_3$}})
+
({\mbox{\boldmath $\sigma_2$}}\cdot{\bf Q})
({\mbox{\boldmath $\sigma_3$}}\cdot{\bf Q})
({\mbox{\boldmath $\tau_2$}}\cdot {\mbox{\boldmath $\tau_3$}})
\over
\omega_\pi^2
}
\right]  \nonumber \\
&&\times \, {
\tilde t_{12}({\bf p},{\bf p'};E-{{q}^2\over 2\nu} - m_\pi)
\over
2m_\pi}
\nonumber \\
&& +{f_{\pi NN}^2(Q)\over m_\pi^2}{1\over (2\pi)^3}
\, {
\tilde t_{12}({\bf p},{\bf p}';E-{{q'}^2\over 2\nu} - m_\pi)
\over
2m_\pi}
\nonumber   \\
& & \!\!
 \times \! \left [
{
({\mbox{\boldmath $\sigma_1$}}\cdot{\bf Q})
({\mbox{\boldmath $\sigma_3$}}\cdot{\bf Q})
({\mbox{\boldmath $\tau_1$}}\cdot {\mbox{\boldmath $\tau_3$}})
+
({\mbox{\boldmath $\sigma_2$}}\cdot{\bf Q})
({\mbox{\boldmath $\sigma_3$}}\cdot{\bf Q})
({\mbox{\boldmath $\tau_2$}}\cdot {\mbox{\boldmath $\tau_3$}})
\over
\omega_\pi^2
}
\right ] \, . \nonumber
\end{eqnarray}
The momenta ${\bf p,q}$ represent respectively the Jacobi
coordinates of the pair ``$1$$2$'', and spectator ``$3$'', while
$E$ is the 3$N$ energy.  The pion-nucleon coupling constant is
selected by the underlying 2$N$ potential that is considered; for
instance, for the Paris and Bonn potentials, we have used the
``traditional'' value $f_{\pi NN}^2/(4\pi)$~=~0.078, while for the
newer CD Bonn potential (\cite{Machleidt2001}) we have
consistently used the more recent determinations by the Nijmegen
\cite{Nijmegen} and VPI \cite{VPI} group. Same considerations have
been applied for the pion-nucleon form factor, since we have
employed the same standard functions (and also with the same
parameters) that have been employed at the level of the 2$N$
potentials:
\begin{equation}
f_{\pi NN}(Q)=f_{\pi NN}{\Lambda_\pi^2-m_\pi^2 \over \Lambda_\pi^2 +Q^2} \, .
\end{equation}
The transferred momentum ${\bf Q}={\bf q'}-{\bf q}$ enters
also in $\omega_\pi=\sqrt{m_\pi^2+Q^2}$.
$\tilde t_{ij}$ denotes the subtracted $t$ matrix between nucleons $1$ and $2$,
defined according to the prescription
\begin{equation}
\tilde t_{12}({\bf p},{\bf p'};Z)=c(E) \, t_{12}({\bf p},{\bf
p'};Z) -v_{12}({\bf p},{\bf p'}) \, . \label{SUB-1}
\end{equation}

\vspace{-2mm}

We have considered in addition another possible type of subtraction
\begin{equation}
\tilde t_{12}({\bf p},{\bf p'};Z)=c(E)\, t_{12}({\bf p},{\bf
p'};Z) -t_{12}({\bf p},{\bf p'};-\tilde\Lambda) \, , \label{SUB-2}
\end{equation}
where the subtraction parameter $\tilde\Lambda$ has been fixed around
1.5 GeV.  Other details can be found in
Refs.\cite{Canton00,Canton01-b}.  The factor $c(Z)$ is an adjustable
parameter and serves to control the cancellation between the two
terms.  Ideally, this factor should be approximately one if the 2$N$
$t$ matrix could be reliably extrapolated off-shell down to $Z\simeq
-160$ MeV. However, the existing 2$N$ potentials cannot guarantee the
extrapolation at such negative energies. Moreover, additional
approximations and simplifications entered in the determination of the
expression for $V^{3N}_3$, as discussed in
Ref.~\cite{Canton00,Canton01-b}.  For these reasons, we introduced
$c(Z)$ as adjustable parameter, with the constrain that at higher
energies in $nd$ scattering, this factor should move towards one,
because the 2$N$ $t$ matrix entering in $V^{3N}_3$ is then calculated
at energies higher than $\simeq -160$ MeV, that is, less far off
shell.  The use of the second type of subtraction, Eq.~\ref{SUB-2},
has been introduced in the case of the Paris potential because this
potential is not OBE-like (one-boson-exchange).  In this case, the
subtraction of the meson-exchange diagrams involved in the
cancellation is therefore not feasible with Eq.~\ref{SUB-1}. On the
other hand Eq.~\ref{SUB-2} tends to enhance the cancellation effects
with respect to Eq.~\ref{SUB-1}, and therefore it is to be expected
that the parameter governing the cancellation in Eq.~\ref{SUB-2} has
to compensate this effect.

The irreducible pionic effect described by Eq.~\ref{OPE3NF} can be
incorporated in the scattering equation in a convenient way if the
$2N$ input potential is of finite rank.  For a rank one (separable)
case, the $2N$ $t$ matrix takes the expression $t=|g_1\rangle \tau
\langle g_1|$, and the (anti)symmetrized AGS equation can be
reinterpreted as an effective two-body multichannel integral equation
in one intercluster momentum variable,
\begin{equation}
X_{11}=Z_{11}+Z_{11}\tau X_{11},
\end{equation}
with the driving term calculated as follows
\begin{equation}
Z_{11}= \langle g_1| G_0 P |g_1\rangle +
\langle g_1| G_0 V^{3N}_1 G_0 |g_1\rangle \, .
\end{equation}

\noindent
The first contribution represents the standard AGS driving term, with
$G_0$ and $P$ being the free Green's function and the
cyclic/anti-cyclic permutator, respectively, while the second
expression takes into account the effects of the irreducible $3N$
diagram discussed above.  The procedure is consistent with the
formalism developed in Ref.\cite{Canton01-a} to include irreducible
pionic effects as corrections in the Faddeev-AGS equation. Since the
formalism is based on the systematic 4-body approach of
Ref.~\cite{Canton98}, it allows to include additional (high-order)
classes of irreducible diagrams in subsequent steps.

Neutron-deuteron scattering observables below 20 MeV have been
calculated using the finite-rank representation of realistic
nucleon-nucleon potentials, known as PEST, BBEST, and CDBEST
potentials \cite{Koike87,Graz,Haidenbauerprivate}.  These provide
accurate representation of the nucleon-nucleon transition matrix for
the Paris~\cite{Paris}, Bonn $B$~\cite{Bonn}, and
CD-Bonn~\cite{Machleidt2001} potentials and are based upon the
Ernst-Shakin-Thaler (EST) method~\cite{EST} for generating the
finite-rank expressions of the transition matrices and/or potentials.
Benchmark calculations \cite{Cornelius90,Schadow98} for scattering and
bound-state regime have demonstrated that with this method it is
possible to solve accurately the Faddeev-AGS scattering equations, and
to obtain results comparable (with errors of 1\% or less) to those
obtained from a direct solution of the 2-dimensional Faddeev equations
(where the original 2$N$ potential is used as input).  The results
shown in the next section have been calculated with the 2$N$
potentials acting in the $j\le 2$ states, and listed in
Tab.~\ref{tab1}. For each 2$N$ potential and state, the table also
reports the rank of the separable expansion used.

Finally, the calculations reported herein have been performed
including all 3$N$ states with total angular momenta up to $J = 19/2$,
for both odd and even parities.

\section{results}
\label{Results}

\begin{figure}
\centerline{\hbox{
\psfig{figure=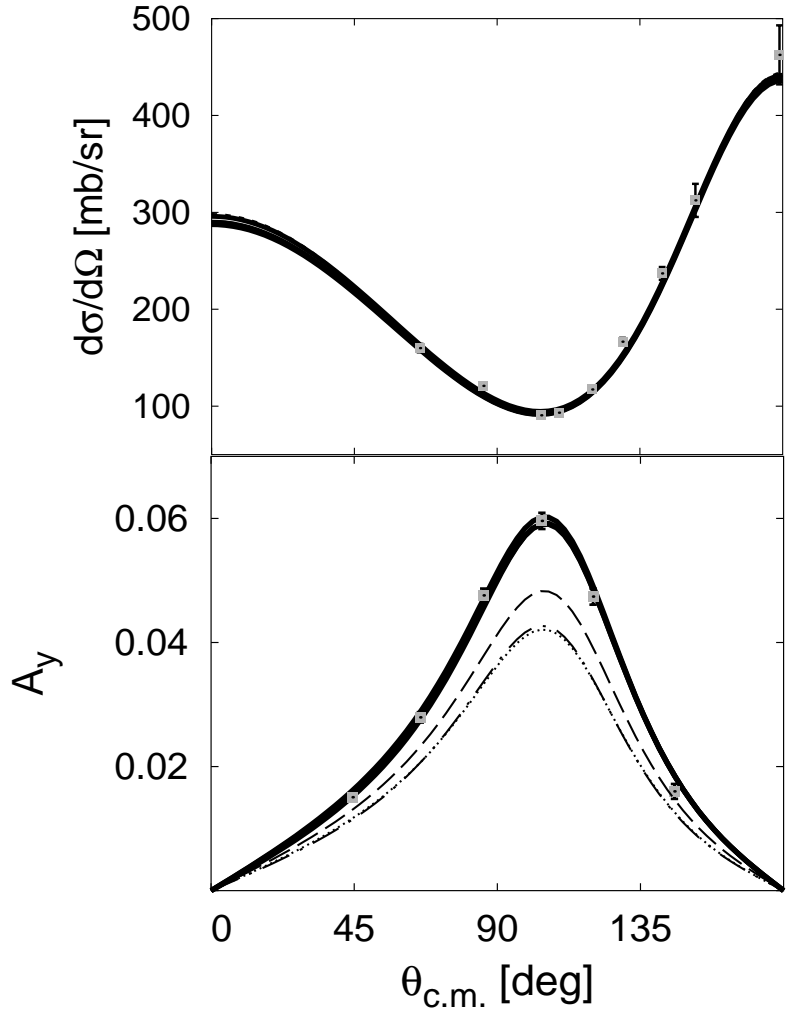,width=132mm,angle=-90}
}}
\vspace{2mm}
\caption[ ]{Differential cross-section and analyzing
power for  $nd$ scattering at 3 MeV (Lab).
Calculations with the PEST potential (dotted line),
BBEST (dot-dashed) and CD-BEST (dashed). For each 2$N$ potential,
the corresponding ranks are given in Tab.\ref{tab1}.
The thick solid line contains the resulting modifications
introduced by the OPE-3NF effect, for all three potentials.
Data (grey squares) from Refs.~\cite{Schwarz83} and \cite{McAninch94}.}
\label{FIG-dcr-Ay-3}
\end{figure}

We first compare the results of our calculations with experimental
data taken with incident neutrons at 3 MeV. At this energy the process
is below the deuteron break-up threshold.  The experimental data shown
in Fig.~\ref{FIG-dcr-Ay-3} are taken from Refs.~\cite{Schwarz83} and
\cite{McAninch94} for the differential cross-section and the neutron
analyzing power, respectively.  The figure compares the calculations
obtained with the CD-Bonn, Bonn $B$, and Paris potential (dashed,
dot-dashed, and dotted lines, respectively) with those obtained when
adding consistently the OPE-3NF effect as discussed in the previous
section.  By adjusting the parameter in Eq.~\ref{SUB-1} (or in
Eq.~\ref{SUB-2} for the Paris potential), we obtain for each
considered potential the required modifications for the proper
reproduction of the analyzing power without ruining the description of
the differential cross section.  For the Bonn $B$ and CD-Bonn potentials
the actual values of the parameter have been reported in
Tab.~\ref{Parameters}. (For the Paris case the parameter has been set
to 0.385).

For all three potentials the result of the calculations including
the OPE-3NF effect is contained in the thick solid line. For the
differential cross-section, there is a very tiny effect or
tendency to reduce the differential cross-section in both the
forward and backward direction, however, such an effect can hardly
be perceived in the figure.

Then, we compare our results with other polarization observables,
taken at comparable energies.  To do so, we consider the complete set
of deuteron analyzing power measurements in deuteron-proton scattering
at 8 MeV, Ref.~\cite{shimizu95}.  The measurements at this energy (for
incident deuterons) compares to an equivalent energy of 4 MeV for the
case of incident protons. However, we do not include Coulomb
corrections in our calculations. As minimal Coulomb correction, we
consider exclusively the additional loss of kinetic energy of the
proton while approaching the electric field of the
deuteron~\cite{Tornow92}.  This loss amounts to about $\simeq 0.7$ MeV
and hence we compare the experimental data of Ref.~\cite{shimizu95}
with theoretical results obtained with incident neutrons of 3.3 MeV.
We observe that the Coulomb slow-down effect has been observed
experimentally for $A_y$ and not for the deuteron analyzing powers,
since there are no experimental neutron data here. However, a recent
calculation~\cite{Kievsky01-a} (with and without Coulomb force)
suggests that the same effect holds also for these observables.

\begin{figure}
\centerline{\hbox{
\psfig{figure=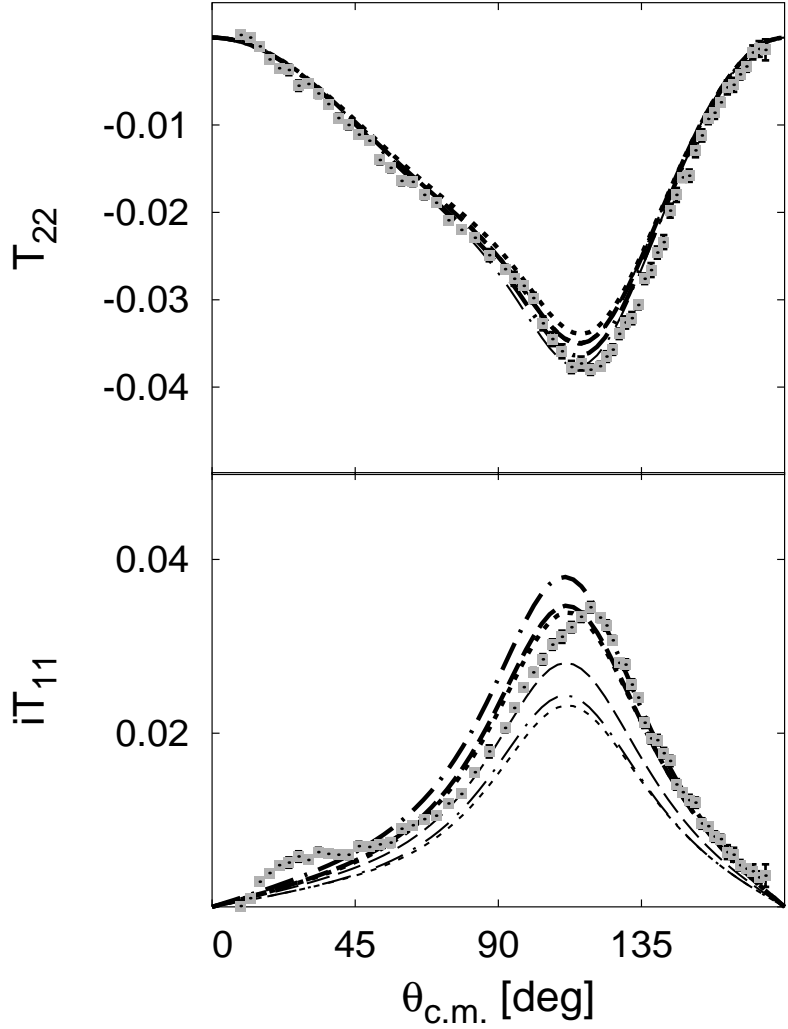,width=132mm,angle=-90}
}}
\vspace{2mm}
\caption[ ]{Deuteron analyzing powers $T_{22}$ (upper panel)
and $iT_{11}$ (lower panel) at 3.3 MeV.
The three thin lines are calculations with 2$N$ potentials only.
The corresponding thick lines contain also the OPE-3NF effects.
As in the previous figure, dotted, dot-dashed, and dashed lines
represent respectively the results with PEST, BBEST, and CD-BEST
potentials. Data are from Ref.~\cite{shimizu95}.}
\label{FIG-T22-iT11-3.3}
\end{figure}

The lower panel of Fig. \ref{FIG-T22-iT11-3.3} considers the deuteron
vector analyzing power $iT_{11}$. Again, we observe that the standard
two-nucleon calculations without 3$N$$F$ effects (thin lines)
underpredict this observable, while the inclusion of the OPE-3NF
effect provides the necessary modifications (thick lines) suggested by
the data.  Conventions for the lines are the same as in the previous
figure, namely, dashed, dot-dashed, and dotted lines, describe
calculations with CD-Bonn, Bonn $B$, and Paris Potentials,
respectively. We observe that the inclusion of the OPE-3NF effect for
the Bonn $B$ potential provides the largest effect for $iT_{11}$, while
for the two other potentials the results with the 3$N$$F$ effect are
rather similar.  This is at variance with respect to the case without
3$N$$F$ effects, where the CD-Bonn potential alone provides a
substantially higher $iT_{11}$ with respect to the Paris and Bonn $B$
cases. Aside for the substantial increase of this observable provided
by the OPE-3NF effect, it is difficult to draw any additional
conclusions by comparison with experimental data because of the
presence of the perturbation introduced by the Coulomb field, which at
these energies modifies sensibly the shape of this observable not only
in the most forward direction.

\begin{figure}
\centerline{\hbox{
\psfig{figure=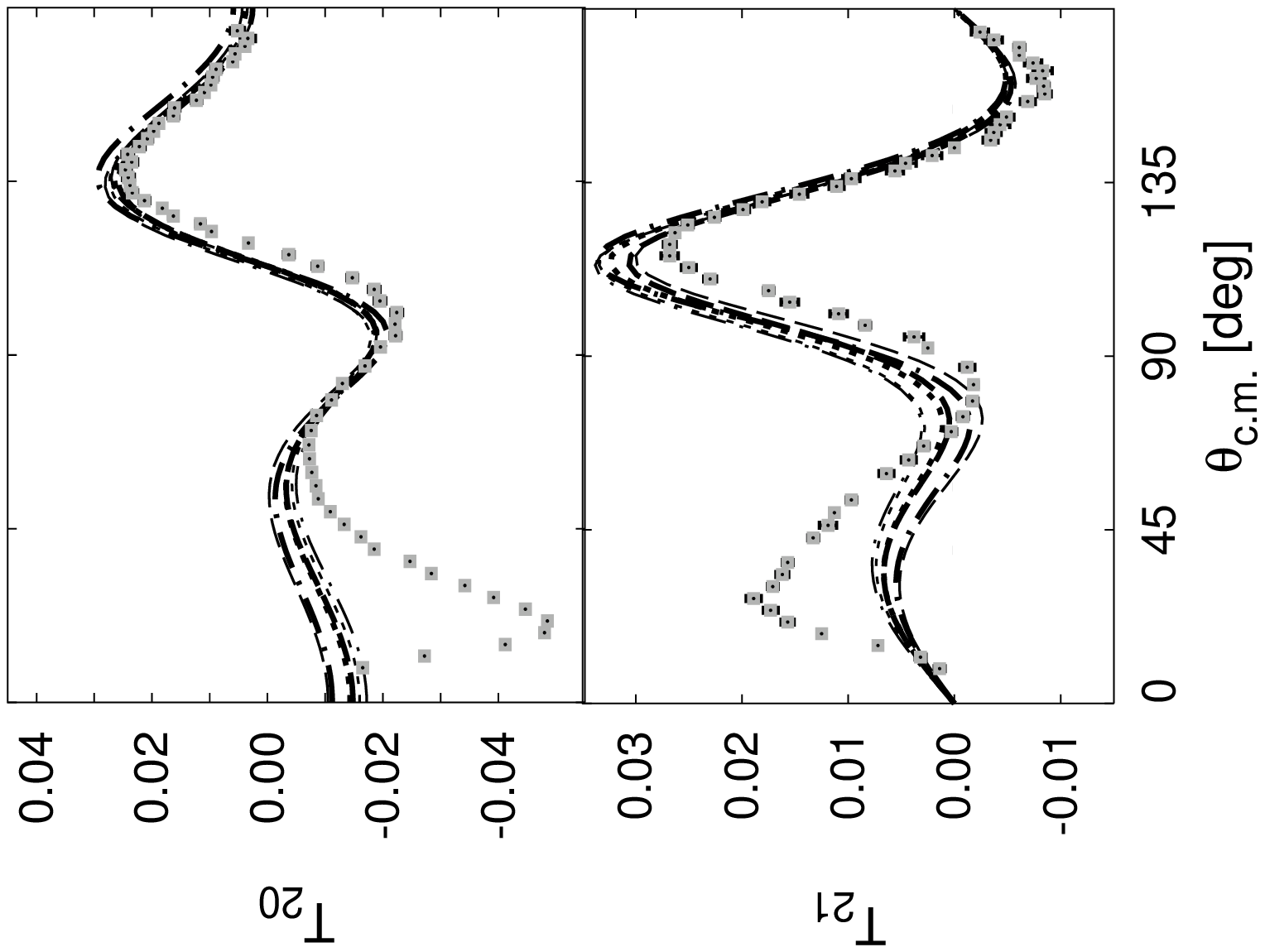,width=132mm,angle=-90}
}}
\vspace{2mm}
\caption[ ]{
Same as in Fig.~\ref{FIG-T22-iT11-3.3}, but for
the $T_{20}$ and $T_{21}$ analyzing powers.
}
\label{FIG-T20-T21-3.3}
\end{figure}

The upper panel of Fig.~\ref{FIG-T22-iT11-3.3} considers the
deuteron tensor analyzing power, $T_{22}$. Here the modification
introduced by the OPE-3NF are very small and cannot be perceived
in the figure, with the exception of the Bonn $B$ case, where a
slight reduction of the dip peaked around 100$^o$ can be observed.
A similar situation is suggested for the other two deuteron tensor
analyzing powers, $T_{20}$ and $T_{21}$, shown respectively in the
upper and lower panel of Fig.~\ref{FIG-T20-T21-3.3}.  The
introduction of the OPE-3NF effect introduces very slight
modifications also in these two observables. Again we observe that
with the inclusion of this 3$N$$F$ effect the Paris and CD-Bonn
results become more similar than they were before. Also with the
Bonn $B$ potential the modifications are small, although one can
clearly perceive for $T_{20}$ that the dip at 100$^o$ and the peak
in the backward direction are slightly more pronounced. Same
situation occurs for the dip at $80^o$ for $T_{21}$. Comparison
with data shows that at these energies and for these observables
no definite conclusions can be drawn without an accurate inclusion
of the Coulomb field, or without a comparison with accurate
measurements involving neutron-deuteron scattering. The figure
also suggests that with proper inclusion of Coulomb modifications
in the theory and/or accurate neutron-deuteron measurements one
could actually observe the phenomenological effects due to this
3$N$$F$-like contribution.

\begin{figure}
\centerline{\hbox{
\psfig{figure=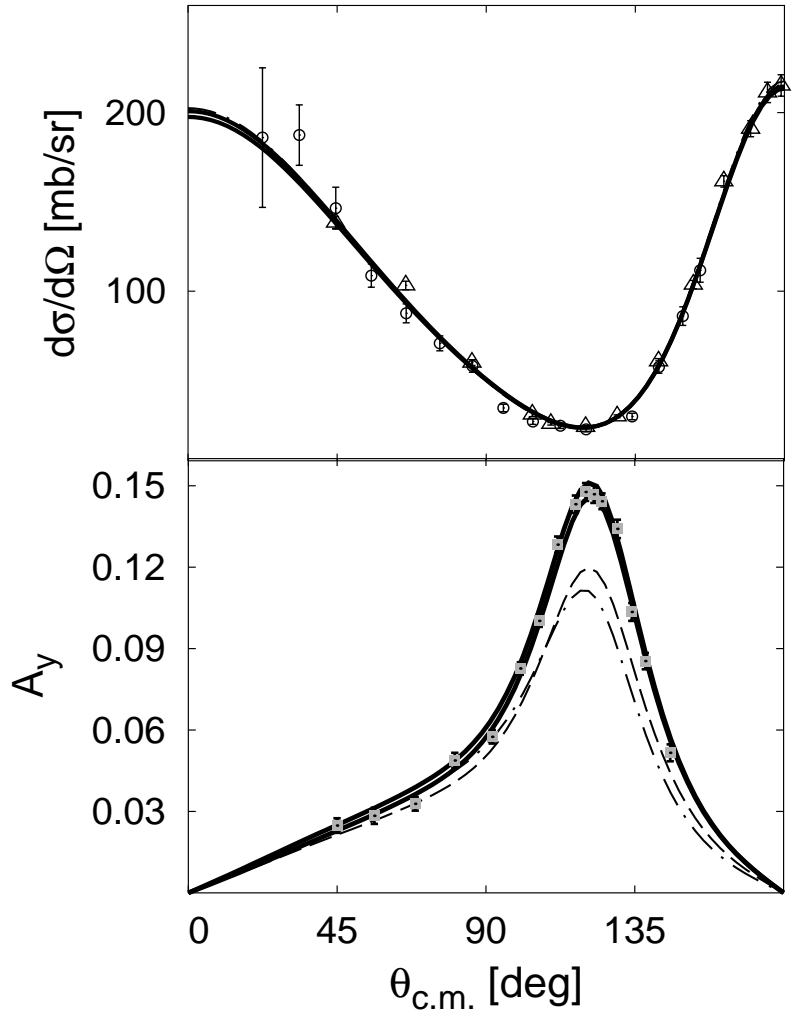,width=132mm,angle=-90}
}}
\vspace{2mm}
\caption[ ]{Same as in Fig.~\ref{FIG-dcr-Ay-3}
but for $n$$d$ scattering at 8.5 MeV.
Data are from Refs.~\cite{Schwarz83},~\cite{Hongquin86}, and
\cite{Tornow91}.
Calculations are for CD-BEST (dashed line) and BBEST (dash-dotted line)
potentials. The thick solid line includes the irreducible 3$N$$F$-like
effects for both potentials.
}
\label{FIG-dcr-Ay-8.5}
\end{figure}

We have repeated the same analysis at 8.5 MeV.
Fig.~\ref{FIG-dcr-Ay-8.5} exhibits that basically the same description
found at 3 MeV holds also at this energy.  The 3$N$$F$ mechanism under
scrutiny is capable to raise the neutron-deuteron $A_y$ in order to
match the experimental data without any sensible modification of the
corresponding differential cross section.  As a minor effect, also at
this energy we can observe a reduction of the cross-section in the
forward and backward directions, for both Bonn $B$ and CD-Bonn
potentials, but in both cases the effect is hardly perceptible in the
figure. The cross-section data have been extracted from
Ref.~\cite{Hongquin86} (open circles) and Ref.~\cite{Schwarz83} (open
triangles) while the polarization data have been taken from
Ref.~\cite{Tornow91}.

\begin{figure}
\centerline{\hbox{
\psfig{figure=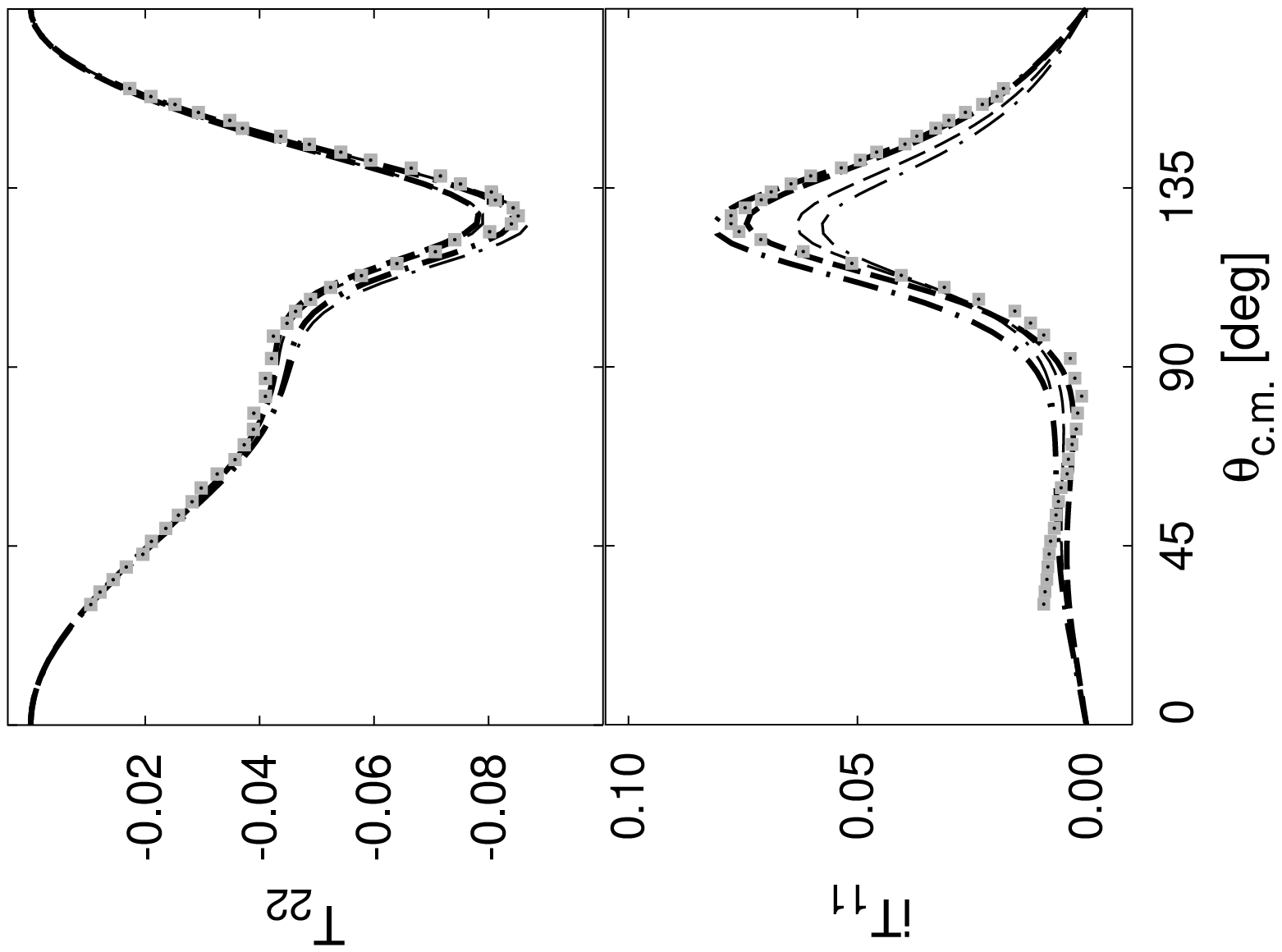,width=132mm,angle=-90}
}}
\vspace{2mm}
\caption[ ]{Same as in Fig.~\ref{FIG-T22-iT11-3.3}, but
for $n$$d$ scattering at 8.3 MeV.
Data are for $p$$d$ scattering at 9 MeV, from Ref.~\cite{Sagara00}.
Thick (thin) lines are calculations with (without) inclusion
of the irreducible pionic effects, as discussed in this work.
The calculations are for the CD-BEST (dashed lines) and BBEST
(dot-dashed lines).
}
\label{FIG-T22-iT11-8.3}
\end{figure}

\begin{figure}
\centerline{\hbox{
\psfig{figure=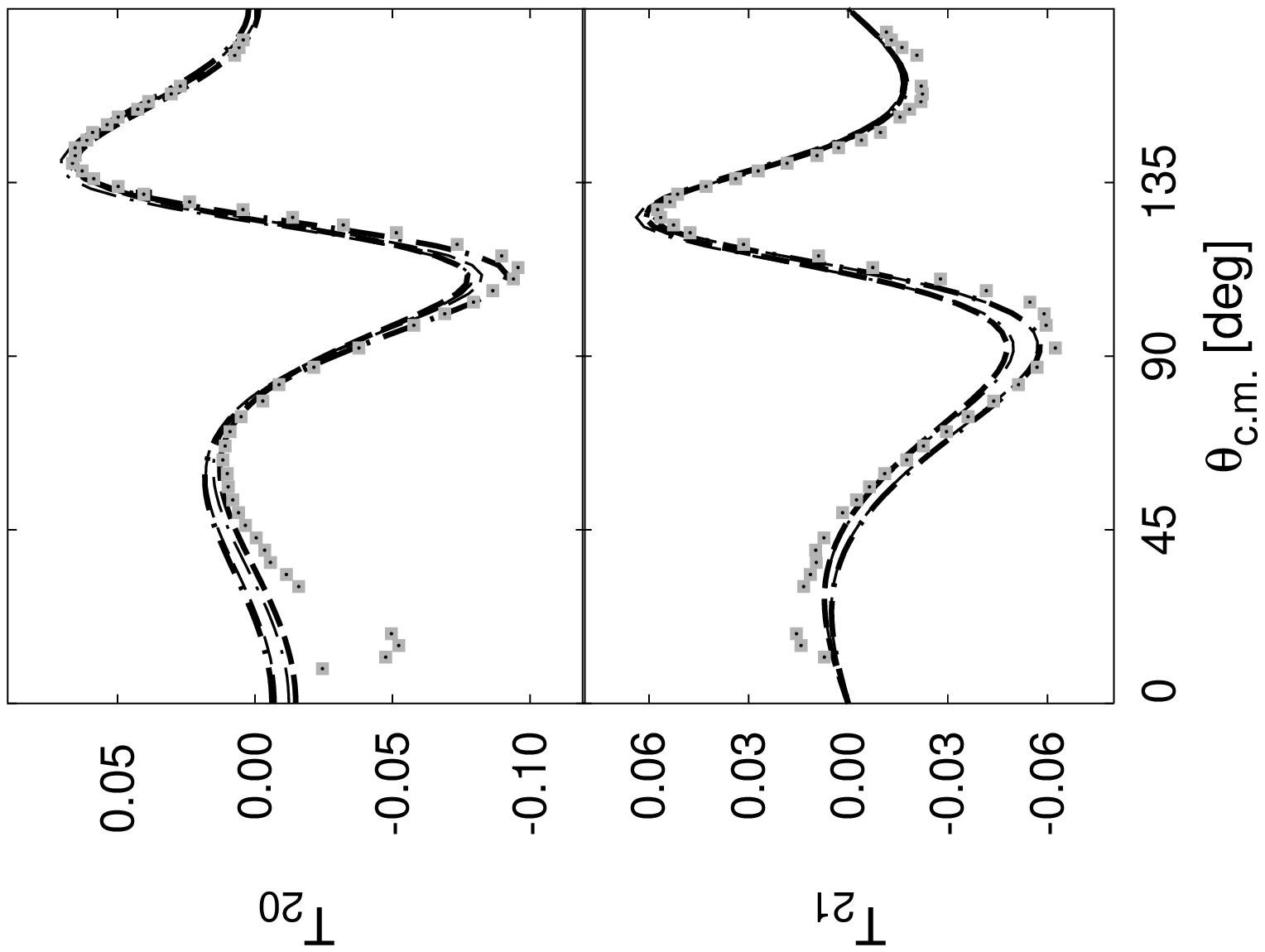,width=132mm,angle=-90}
}}
\vspace{2mm}
\caption[ ]{
Same as in Fig.~\ref{FIG-T22-iT11-8.3}, but for
the $T_{20}$ and $T_{21}$ analyzing powers.
}
\label{FIG-T20-T21-8.3}
\end{figure}

In Figs.~\ref{FIG-T22-iT11-8.3} and \ref{FIG-T20-T21-8.3} we have
compared ``$n$$d$'' theoretical results with ``$p$$d$'' data for the
deuteron analyzing powers.  For the reasons explained above we
compared results obtained at 8.3 MeV (for incident neutron energy)
with data taken at 18 MeV (incident deuterons) \cite{Sagara00}.  The
lower panel of Fig.~\ref{FIG-T22-iT11-8.3} shows that when including
the 3$N$$F$ effect the increase of $iT_{11}$ is about of the right size
for both potentials; however one observes also an inversion of the
peaks, since, when including the 3$N$$F$ effect, the Bonn $B$ peak is
higher than the CD-Bonn one, while without 3$N$$F$ the CD-Bonn peak is
higher.  The same thing was occurring also at lower energy.

For the deuteron tensor analyzing powers, $T_{22}$ (upper panel of
Fig.~\ref{FIG-T22-iT11-8.3}), $T_{20}$ and $T_{21}$ (upper and lower
panels of Fig.~\ref{FIG-T20-T21-8.3}, respectively), we observe that
the overall shape of the angular distributions is better reproduced
than at 4 MeV, indicating that the effects of Coulomb distortions are
less important here (with the exception of the data in the forward
direction), and that the modifications introduced by the 3$N$$F$ effects
are in general of minor entity with respect to those observed for the
vector analyzing powers. A closer inspection, however, reveals that
while the OPE-3NF effects are very small for $T_{22}$ for both CD-Bonn
and Bonn $B$ potentials, the situation is different for the other two
deuteron tensor observables, where the effects of this 3$N$$F$ diagram
can indeed be observed and are about of the same size of the
difference between the two potentials themselves.  For $T_{20}$ in
particular, the 3$N$$F$ effect provides a remarkable improvement of the
dip at $110^o$ in the case of the Bonn $B$ potential, while for the case
of the CD-Bonn potential the situation is basically unchanged. For
$T_{21}$ again the 3$N$$F$ effect remarkably improves the description in
the Bonn $B$ case (especially at the dip around $90^o$), while for the
case of the CD-Bonn potential the situation is reversed.

\begin{figure}
\centerline{\hbox{
\psfig{figure=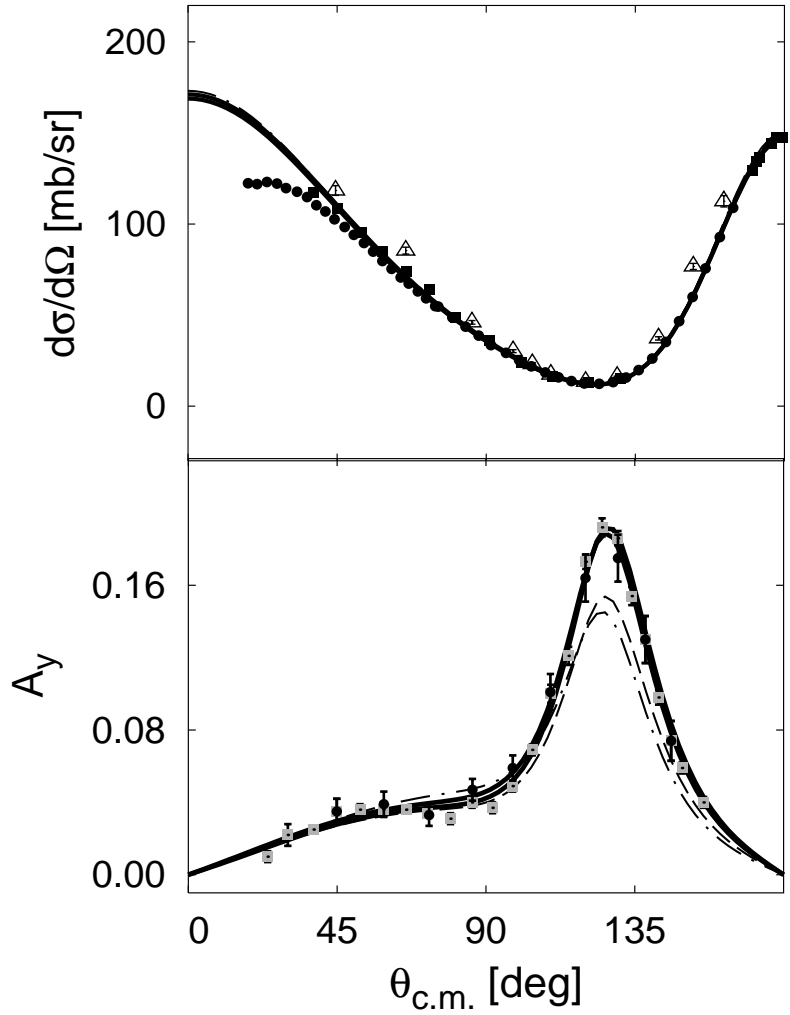,width=132mm,angle=-90}
}}
\vspace{2mm}
\caption[ ]{Same as in Fig.~\ref{FIG-dcr-Ay-8.5}
but for nucleon scattering at 12.0 MeV.
Data for the differential cross-section
are from Refs.~\cite{Schwarz83} (triangles, $n$$d$),~\cite{Sagara94}
(circles, $p$$d$), and \cite{Sydow88} (squares, $p$$d$).
Data for the neutron analyzing power ($A_y$) are from Refs.~\cite{Tornow78}
(black circles) and ~\cite{Tornow89} (gray squares).
Calculations are for CD-BEST (dashed line) and BBEST (dash-dotted line)
potentials. The thick solid line includes the irreducible 3$N$$F$-like
effects for both potentials.
}
\label{FIG-dcr-Ay-12.0}
\end{figure}

\begin{figure}
\centerline{\hbox{
\psfig{figure=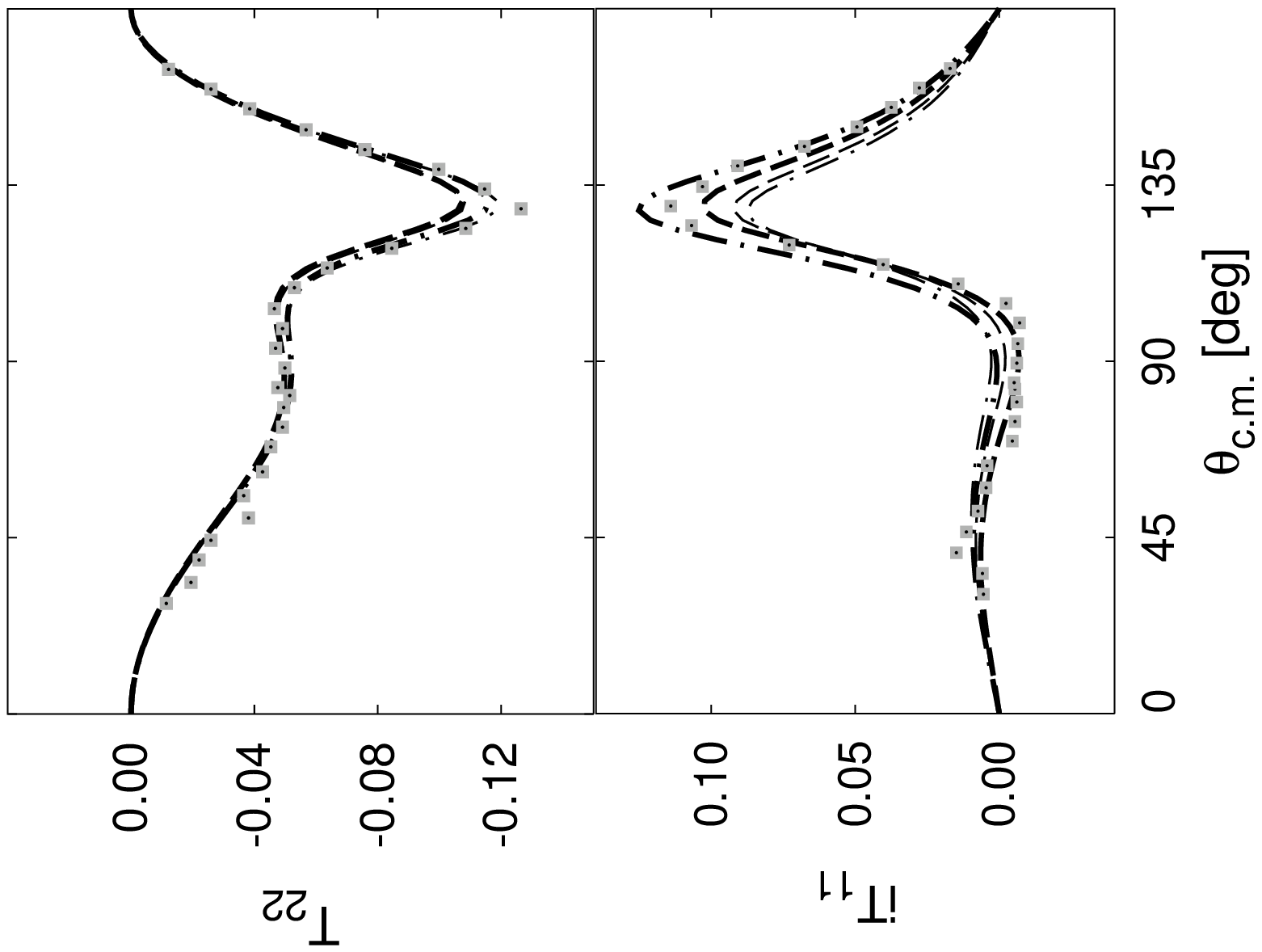,width=132mm,angle=-90}
}}
\vspace{2mm}
\caption[ ]{Same as in Fig.~\ref{FIG-T22-iT11-8.3}, but
for $n$$d$ scattering at 11.3 MeV.
Data are for $p$$d$ scattering at 12 MeV, from Ref.~\cite{Gruebler83}.
Thick (thin) lines are calculations with (without) inclusion
of the irreducible pionic effects, as discussed in this work.
The calculations are for the CD-BEST (dashed lines) and BBEST
(dot-dashed lines).
}
\label{FIG-T22-iT11-12.0}
\end{figure}

\begin{figure}
\centerline{\hbox{
\psfig{figure=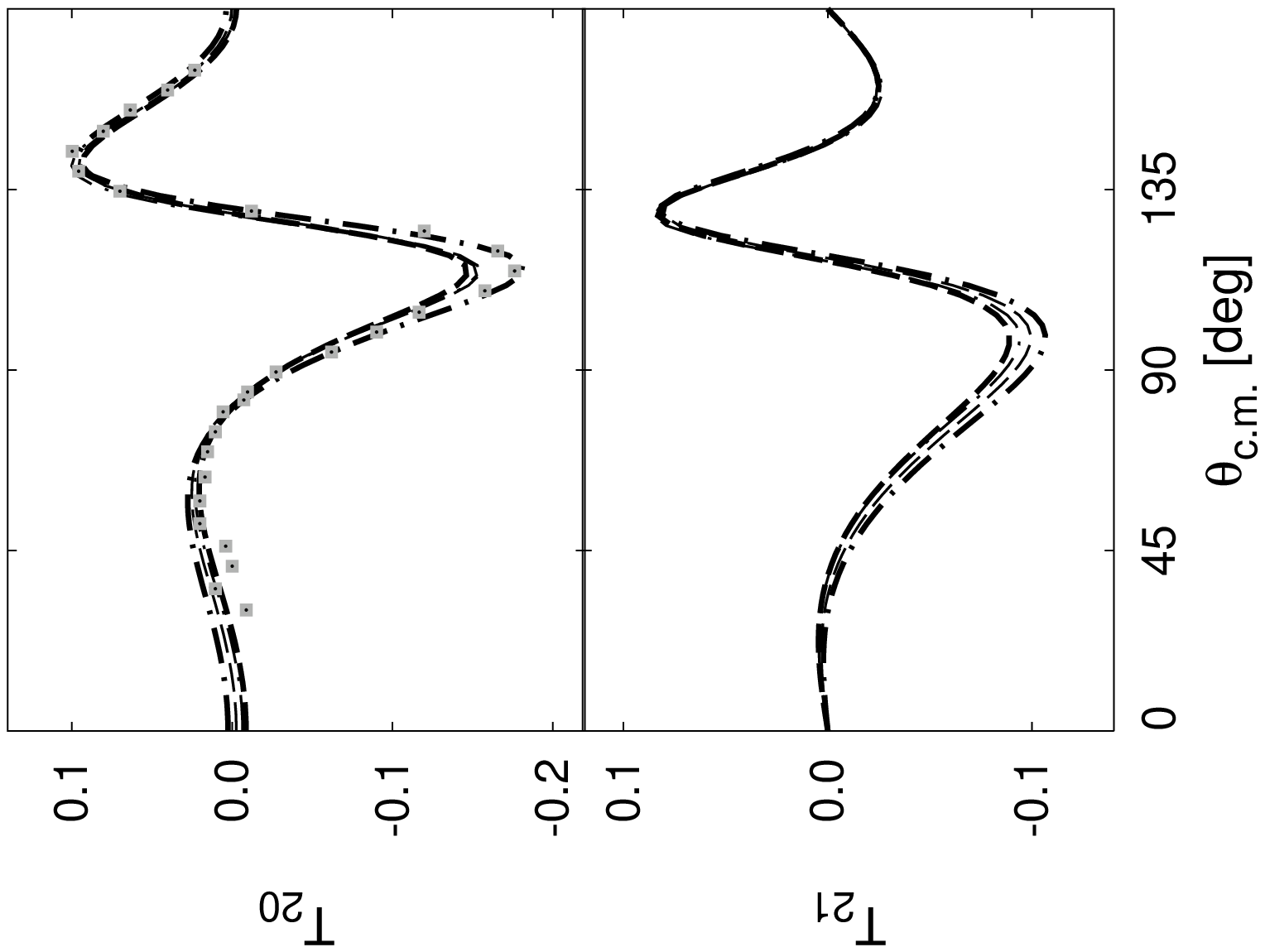,width=132mm,angle=-90}
}}
\vspace{2mm}
\caption[ ]{
Same as in Fig.~\ref{FIG-T22-iT11-12.0}, but for
the $T_{20}$ and $T_{21}$ analyzing powers.
}
\label{FIG-T20-T21-12.0}
\end{figure}

Then, we have considered how the situation evolves at 12 MeV.
Again we find that the introduction of these irreducible pionic
effects are able to increase significantly $A_y$ (lower panel of
Fig.~\ref{FIG-dcr-Ay-12.0}) for both potentials without affecting
appreciably the differential cross-section (upper panel of
Fig.~\ref{FIG-dcr-Ay-12.0}).  As the energy increases, it becomes
evident that the introduction of these 3$N$$F$-like effects
provide a good description of $A_y$ in both hemispheres. To obtain
this, the corrections have to operate differently in the two
directions, with a substantial increase of the analyzing power in
the backward hemisphere, where the peak evolves, and at the same
time with a slight suppression of $A_y$ in the forward hemisphere,
especially for the Bonn $B$ potential. Clearly, these irreducible
pionic effects have the ability to achieve both goals.  In
Figs.~\ref{FIG-T22-iT11-12.0} and \ref{FIG-T20-T21-12.0} we
compare $p$$d$ data measured at the proton-equivalent energy of 12
MeV with $n$$d$ calculations at 11.3 MeV (for consistency with the
Coulomb slow-down assumption). For both potentials, the
3$N$$F$-like effects increase $iT_{11}$ significantly in the
backward hemishere and suppress slightly the observable at forward
angles.  Comparison with experimental data suggest that these
modifications are correct in both directions, although great
caution has to be exercised when comparing $n$$d$ calculation with
$p$$d$ data. Around the peak at backward angles we observe also a
quite large increase for the case of the Bonn $B$ potential, while
the effect is smaller for the CD-Bonn case.  This last feature is
similar to what has been observed at lower energy.

The 3$N$$F$-like effects have a very minor impact on $T_{22}$, also at
12 MeV (upper panel of Fig.~\ref{FIG-T22-iT11-12.0}), while it is
evident that this observable is more sensitive to the choice of the
underlying 2$N$ potential, and in absence of other additional
contributions, we might conclude that the data seem to favor more the
BBEST calculation, with respect to the CD-BEST results.

In Fig.~\ref{FIG-T20-T21-12.0} we consider the remaining two tensor
analyzing powers, $T_{20}$ (upper panel) and $T_{21}$ (lower panel).
$T_{20}$ exhibits an interesting evolution since the results for the
BBEST + OPE-3NF case are significantly different around 110$^o$ than
the other cases, and they are remarkably close to the experimental
data.  An interesting situation occurs also for $T_{21}$ around
100$^o$ where the BEST + OPE-3NF calculation are appreciably more
negative than the other calculations considered in the figure. We
calculated that this same situation evolves also at higher energies
(e.g., 18-20 MeV).  Unfortunately, no experimental data have been
found for $T_{21}$ in this energy range, and new measurements for
$T_{21}$ would be very useful in the range of 10-20 MeV.

As has been explained in Refs.~\cite{Canton00}, one basic aspect
of the irreducible pionic effects which generate the 3$N$$F$
diagram included in our calculation develops as a consequence of
an ``imperfect'' cancellation with respect to the mesonic
retardation contributions.  Since in our study we employ realistic
nucleon-nucleon potentials, we observe that they are heavily based
on fitting procedures of experimental nucleon-nucleon data and are
therefore a sort of ``black boxes'' with respect to the variety
and structure of meson-exchange diagrams included, except probably
the OPE term, since only the longest range of the nuclear force is
well established.  It should therefore not be a surprise that for
each 2$N$ potential we included in the corresponding 3$N$$F$
diagram a phenomenological parameter that governs the level of
cancellation against meson retardation effects.  In the approach
we developed in Ref.~\cite{Canton01-b} we used the experimental
value of $A_y$ at the peak to actually fix this parameter. Then,
it is obviously of interest to study how this parameter evolves
with energy.  For this reason we considered the wealth of $p$$d$
experimental data for $A_y$ measured at 12, 14, 16, and 18 MeV
(Ref.~\cite{Sagara94}) and compared these data with $n$$d$
calculations at 11.3, 13.3, 15.3, and 17.3 MeV, respectively,
finding that we could reproduce how the peak evolves with a
perfectly linear dependence of the parameter governing the
cancellation of the 3$N$$F$.

\begin{figure}
\centerline{\hbox{
\psfig{figure=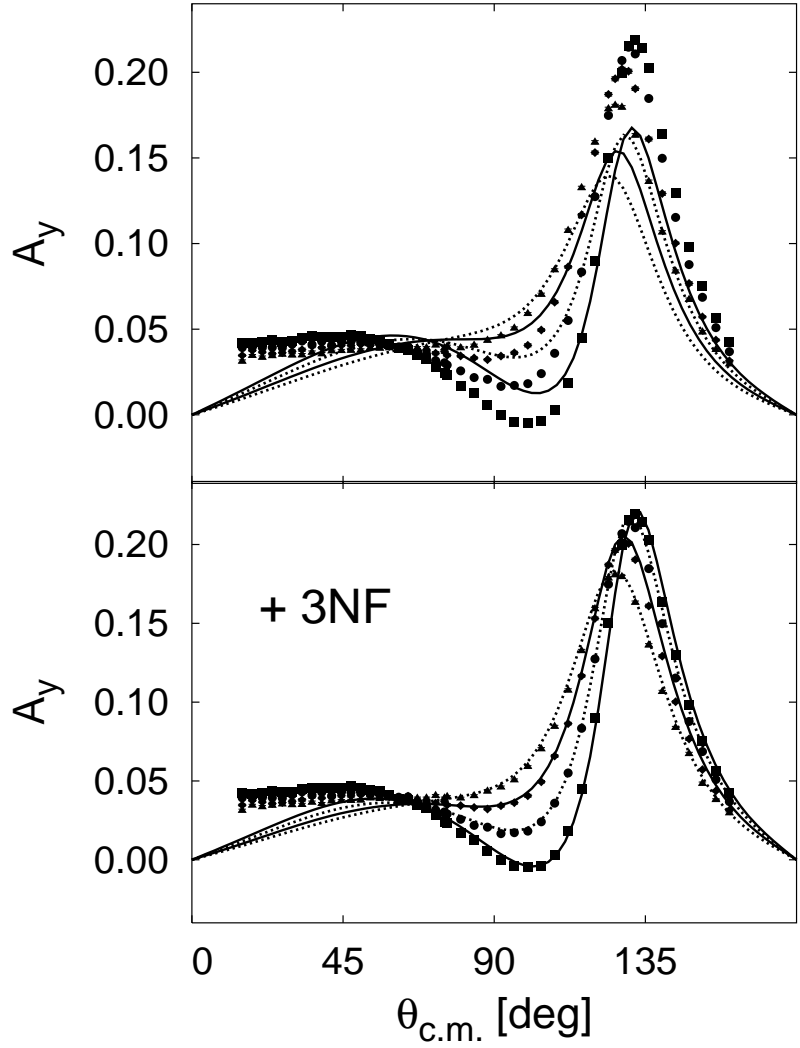,width=112mm,angle=-90}
}} \vspace{2mm} \caption[ ]{Evolution of the analyzing power $A_y$
in the range 12-18 MeV. Squares, circles, diamonds and triangles
represents $p$$d$ data at 18, 16, 14 and 12 MeV, respectively,
taken from Ref.~\cite{Sagara94}. The lines in the upper panel
refer to corresponding calculations with the BBEST potential,
while in the lower panel the OPE-3NF effects are also included. }
\label{BBonn-Ay-12-18}
\end{figure}

\begin{figure}
\centerline{\hbox{
\psfig{figure=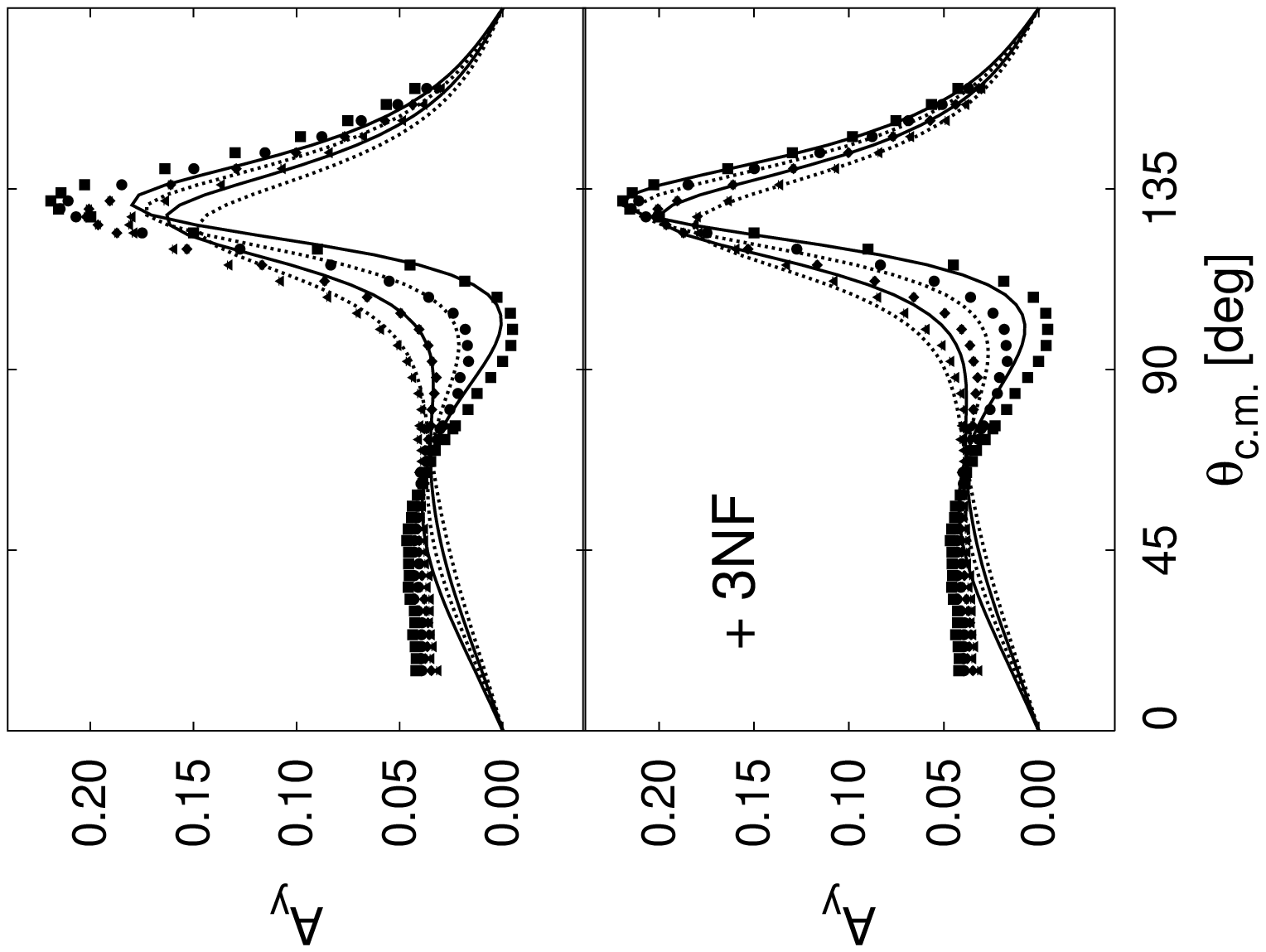,width=112mm,angle=-90}
}} \vspace{2mm} \caption[ ]{Same as in Fig.~\ref{BBonn-Ay-12-18}
but for the CD-BEST potential.} \label{CDBonn-Ay-12-18}
\end{figure}

The situation is shown in Figs.~\ref{BBonn-Ay-12-18} and
~\ref{CDBonn-Ay-12-18} for the Bonn $B$ and CD-Bonn potentials.  In the
two figures, the upper panel compares data with results taken without
3$N$$F$, while the results in the lower panel include 3$N$$F$
effect. Inclusion of the OPE-3NF effect provides very good results for
the Bonn $B$ potential, not only in the region of the peak (around
$130^o$ MeV), but also in the region around 100$^o$ where the dip
evolves. For the CD-Bonn potential the situation is similarly
satisfactorily for the evolution of the $A_y$ peak while the evolution
of the dip at lower angles is fair but not optimal.  For both
potentials, we report in Tab.~\ref{Parameters} the value of the
parameter and the corresponding energy we have employed in calculating
this 3$N$$F$ effect.

In this same energy range, we focus attention on the nucleon-deuteron
polarization transfer coefficients, $K_y^{y'}$, $K_z^{x'}$,
$K_y^{x'x'-y'y'}$, and $K_y^{z'z'}$. It has been suggested in
Ref.~\cite{Sydow94} that these observables are sensitive to the tensor
part of the nuclear forces and therefore they could in principle
represent a good testing ground for the 3$N$$F$-operator we are studying
in the present paper, since this has a pronounced tensor structure.
In addition, recent experimental data are now available, for both $n$$d$
and $p$$d$ systems, and a number of theoretical studies about these
coefficients have been performed
already~\cite{Sydow94,Hempen98,Sydow88,Kievsky01,Sydow98} with a
variety of different 2$N$ potentials, with the addition of 3$N$$F$'s,
and more recently also with the inclusion of the modifications
introduced by Coulomb effects.  From these studies it emerged that the
nucleon-to-nucleon transfer coefficients $K_y^{y'}$ and $K_z^{x'}$
exhibit a scaling behavior with respect to the triton binding energy,
while the nucleon-to-deuteron coefficients $K_y^{x'x'-y'y'}$, and
$K_y^{z'z'}$ do not scale. Moreover, $K_y^{y'}$ and $K_z^{x'}$ exhibit
sizable Coulomb effects, while for the other two coefficients the
effects are much less appreciable.  For the case of $K_y^{y'}$, where
it was possible to compare directly theory with $n$$d$ data, it was found
that the theoretical calculations underpredict the minimum at $110^o$,
once the scaling effect with binding, originated by the
2$\pi$-exchange 3$N$$F$, was properly taken into account. Moreover,
discrepancies between theoretical calculations and experimental data
concerning the Coulomb effects for $K_y^{y'}$ at 19 MeV concur to
conclude that the situation is not fully understood.  For the
nucleon-to-deuteron tensor transfer coefficients the situation is also
unclear, since Coulomb effects and traditional 2$\pi$-exchange 3$N$$F$
provide too small modifications for a correct reproduction of data
(for $K_y^{x'x'-y'y'}$ the peak at $135^o$ is underestimated).  This
state of affairs demands for more theoretical investigations; at the
same time more extensive experimental studies for these observables in
this energy range could be extremely useful.

\begin{figure}
\centerline{\hbox{
\psfig{figure=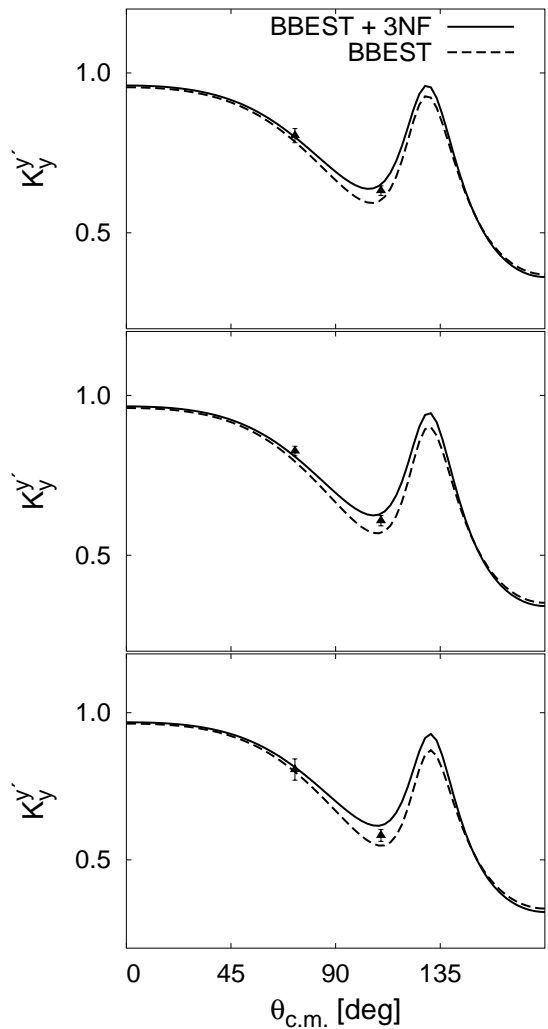,width=112mm,angle=-90}
}}
\vspace{2mm}
\caption[ ]{Spin-transfer coefficient $K_y^{y'}$ at 15 MeV (upper
panel), 17 MeV (middle panel), and 19 MeV (lower panel).  Comparison
between $n$$d$ data from Ref.~\cite{Hempen98} and BBEST calculations.
Solid (dashed) lines include (exclude) the irreducible 3$N$$F$-like
pionic effects.  }
\label{FIG-bbonn-Kyy-15-19}
\end{figure}

\begin{figure}
\centerline{\hbox{
\psfig{figure=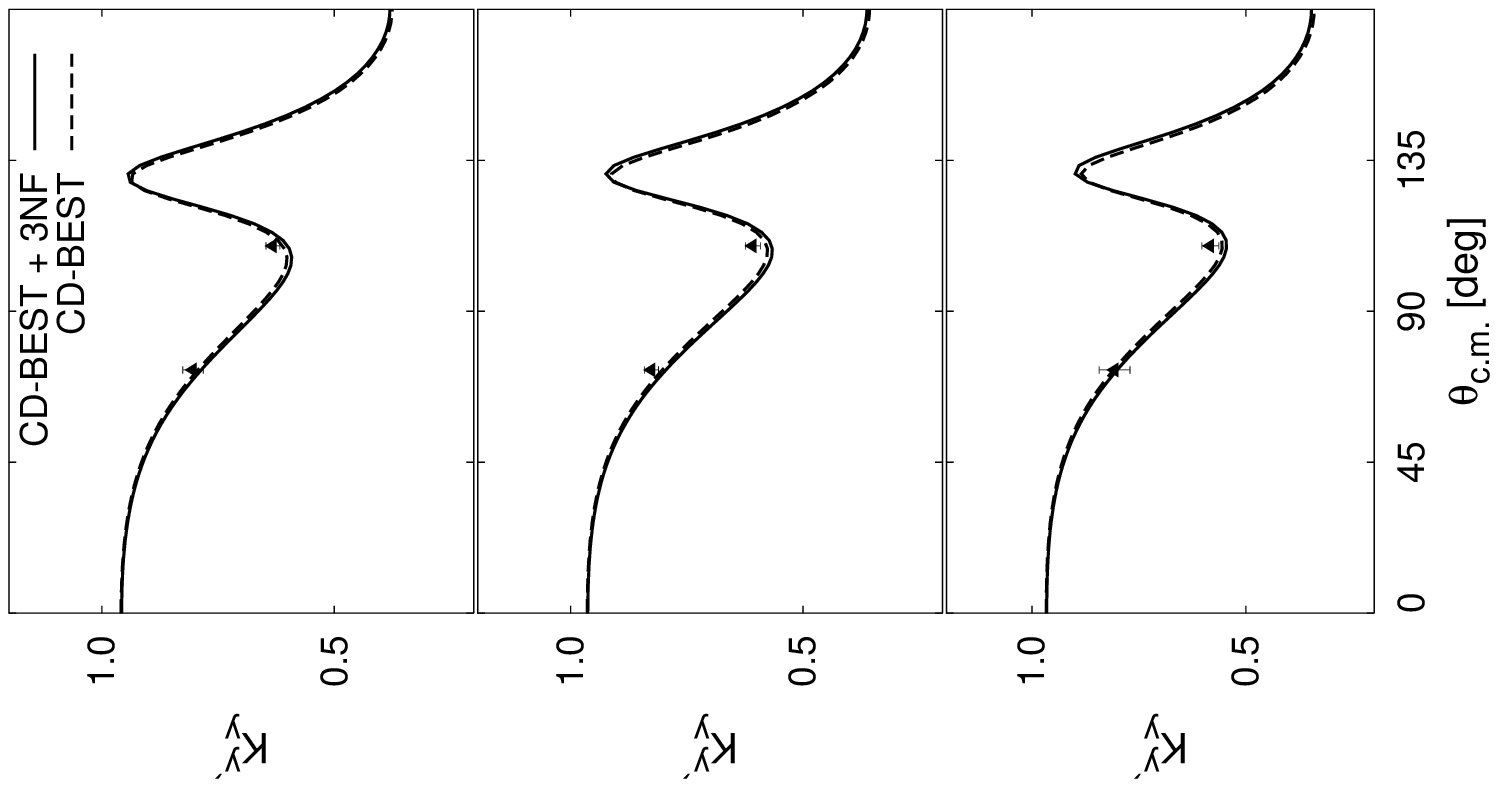,width=112mm,angle=-90}
}}
\vspace{2mm}
\caption[ ]{
Same as in Fig.~\ref{FIG-bbonn-Kyy-15-19}
but for the CD-BEST potential.
}
\label{FIG-cdbonn-Kyy-15-19}
\end{figure}

\begin{figure}
\centerline{\hbox{
\psfig{figure=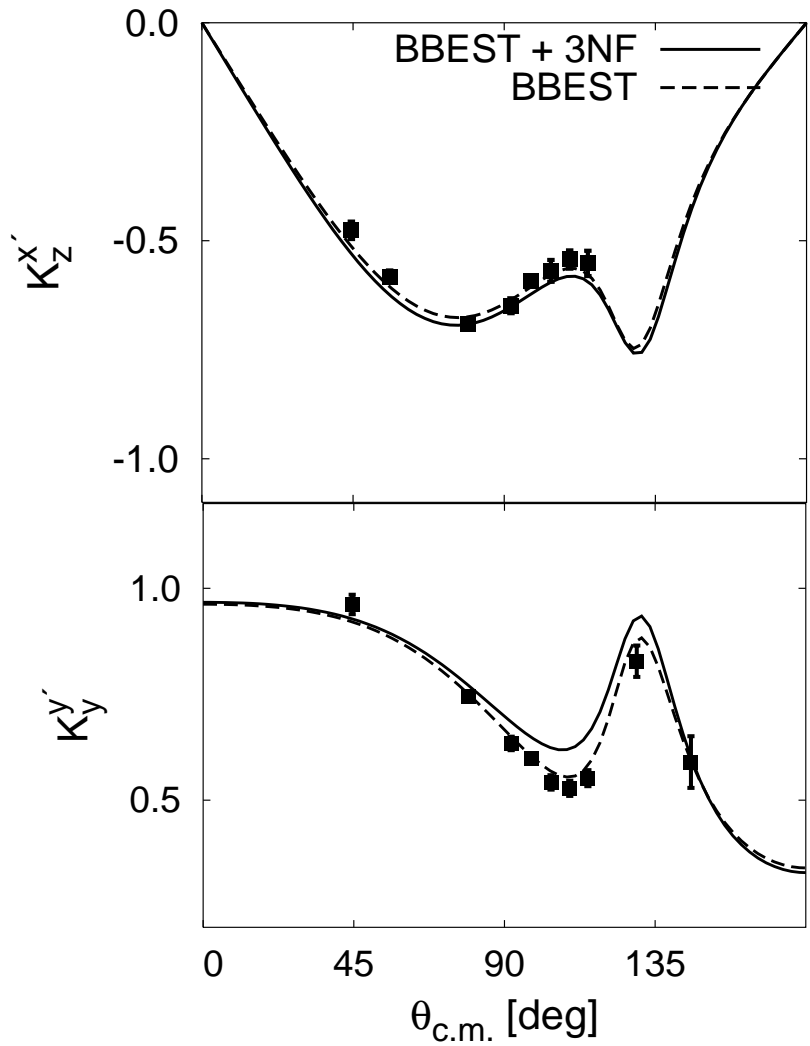,width=112mm,angle=-90}
}}
\vspace{2mm}
\caption[ ]{
Calculations for the $n$$d$ spin-transfer-coefficients $K_z^{x'}$
(upper panel) and  $K_y^{y'}$ (lower panel), at 18.3 MeV,
for the BBEST potential. Data are for the equivalent $p$$d$
observables at 19.0 MeV, from Ref.~\cite{Sydow94}.
Solid (dashed) lines include (exclude) the irreducible 3$N$$F$-like
pionic effects.
}
\label{FIG-pd-bbonn-nnvvK-19}
\end{figure}

\begin{figure}
\centerline{\hbox{
\psfig{figure=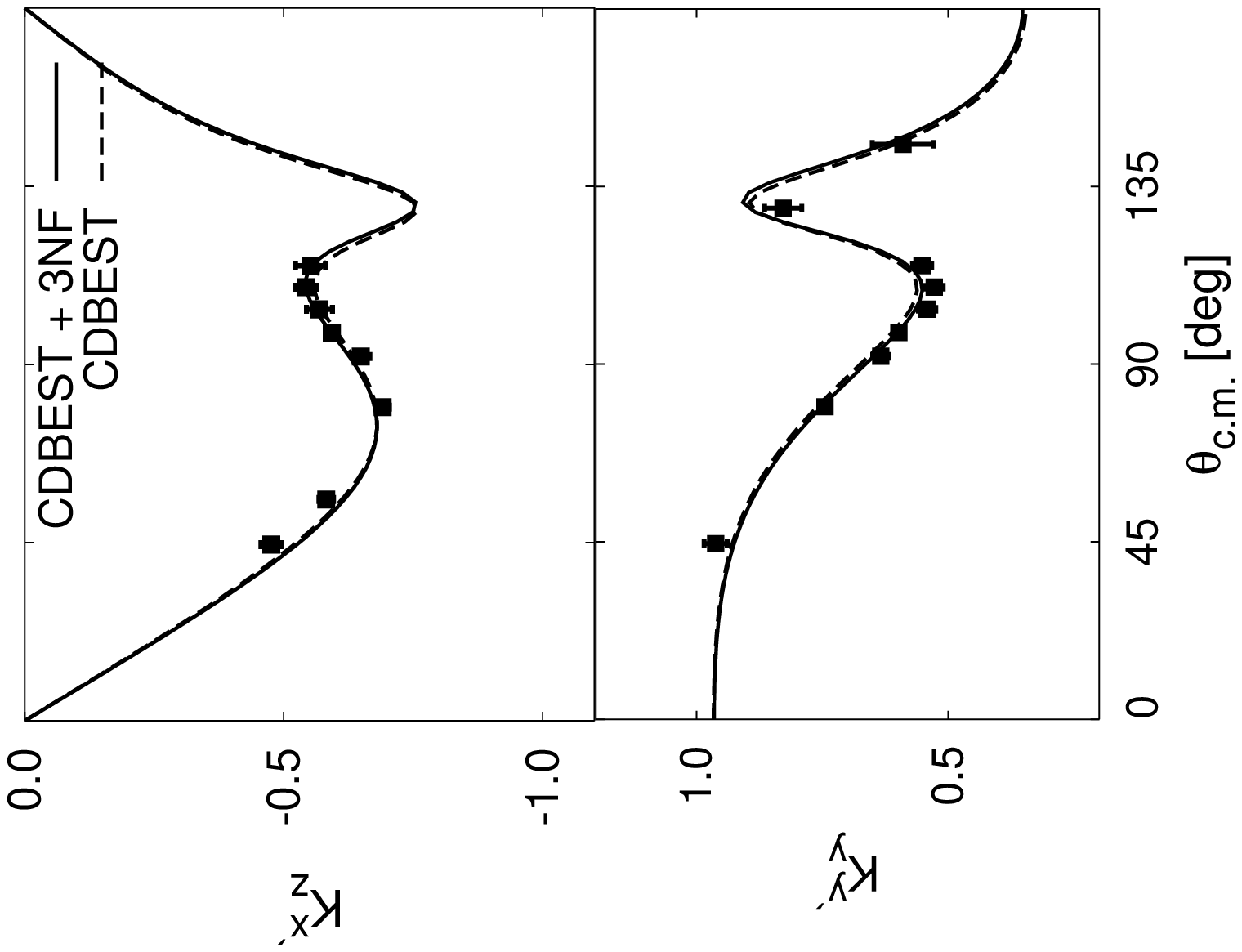,width=112mm,angle=-90}
}}
\vspace{2mm}
\caption[ ]{
Same as in Fig.~\ref{FIG-pd-bbonn-nnvvK-19}
but for the CD-BEST potential.}
\label{FIG-pd-cdbonn-nnvvK-19}
\end{figure}

In Figs.~\ref{FIG-bbonn-Kyy-15-19} and ~\ref{FIG-cdbonn-Kyy-15-19} we
show the results obtained with the Bonn $B$ and CD-Bonn potentials,
respectively, and compare these with experimental $n$$d$ data taken at
15, 17, and 19 MeV, Ref.~\cite{Hempen98}.  Our calculations do not
include the effects of the 2$\pi$-exchange 3$N$$F$, and therefore one
should take into account in the discussion the rescaling effect that
tend to push the lines downward (this tendency is however reduced with
the Bonn-types interactions, which provide smaller 3$N$ underbinding
with respect to other 2$N$ interactions).  The results for the Bonn $B$
case suggest that the 3$N$$F$ effect we have calculated is able to
correct the underprediction of the minimum of $K_y^{y'}$ at $110^o$,
however the results with the CD-Bonn potential do not confirm this
indication, since the modifications are smaller here and have the
tendency to go in the opposite direction.  In
Figs.~\ref{FIG-pd-bbonn-nnvvK-19} and ~\ref{FIG-pd-cdbonn-nnvvK-19} we
similarly compare the $p$$d$ data at 19 MeV \cite{Sydow94} with $n$$d$
calculations at 18.3 MeV. This comparison has to be made with great
prudence, since Coulomb effects and the 2$\pi$-3NF provide appreciable
effects, but the two modifications tend somewhat to cancel
out~\cite{Kievsky01}.  Nevertheless, it is interesting to observe that
here the minimum of $K^{y'}_y$ is appreciably overpredicted in the
Bonn $B$ case (lower panel of Fig.~\ref{FIG-pd-bbonn-nnvvK-19}) while in
the case of the CD-Bonn potential the irreducible pionic effects do
not affect the results appreciably and therefore they maintain the
quality of the fit (lower panel of Fig.~\ref{FIG-pd-cdbonn-nnvvK-19}).
Thus, the experimental ``mismatch'' between the $n$$d$ and $p$$d$ data for
$K_y^{y'}$ at 19 MeV acquires an interesting twist: the $n$$d$ data seem
to support the Bonn $B$ + OPE-3NF calculations, the $p$$d$ data seem to
support more the CD-Bonn + OPE-3NF results.

\begin{figure}
\centerline{\hbox{
\psfig{figure=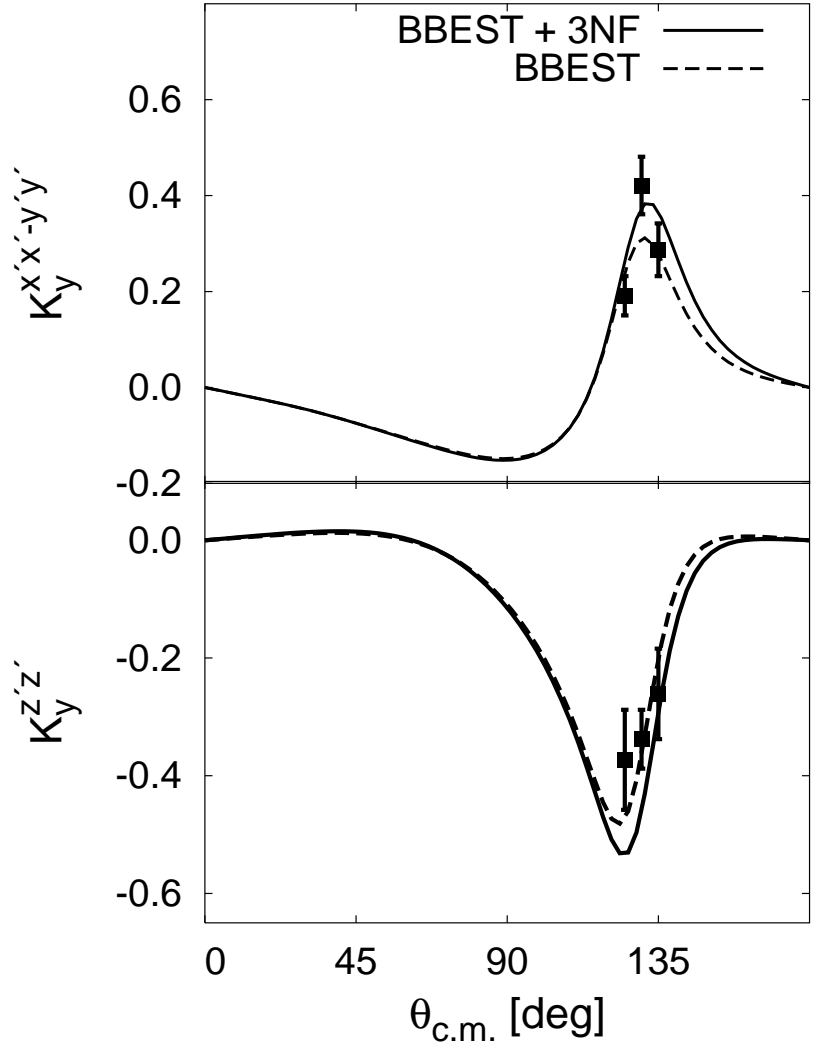,width=112mm,angle=-90}
}}
\vspace{2mm}
\caption[ ]{Polarization transfer coefficients
$K_y^{x'x'-y'y'}$ (upper panel) and $K_y^{z'z'}$ (lower panel) calculated
at 18.3 MeV for the BBEST potential. Data are for $p$$d$ scattering at 19.0 MeV,
from Ref.~\cite{Sydow98}.}
\label{FIG-pd-bbonn-ndvtK-19}
\end{figure}

\begin{figure}
\centerline{\hbox{
\psfig{figure=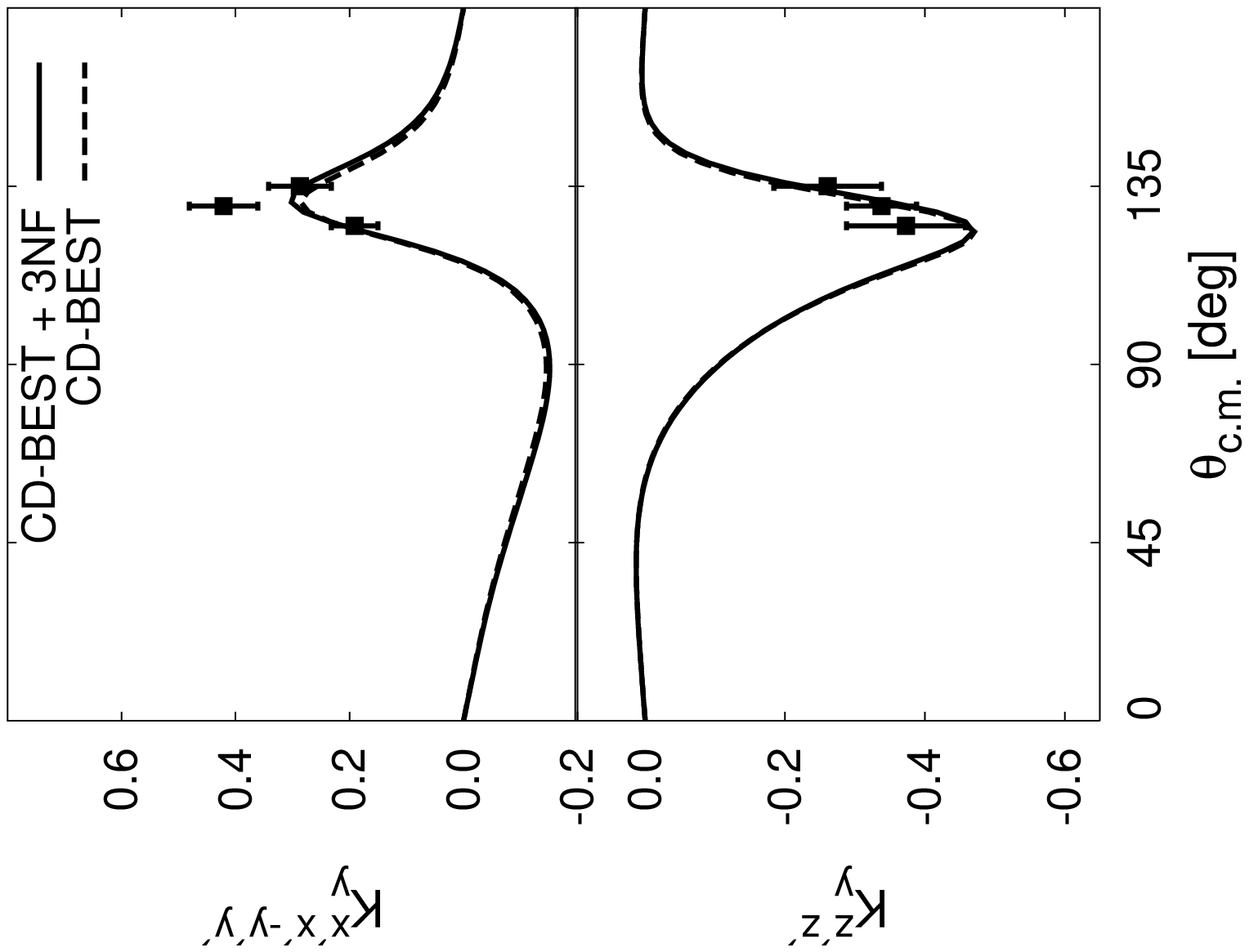,width=112mm,angle=-90}
}}
\vspace{2mm}
\caption[ ]{
Same as in Fig.~\ref{FIG-pd-bbonn-ndvtK-19}
but for the CD-BEST potential.}
\label{FIG-pd-cdbonn-ndvtK-19}
\end{figure}

Finally, in Figs.~\ref{FIG-pd-bbonn-ndvtK-19} and
~\ref{FIG-pd-cdbonn-ndvtK-19} the tensor transfer coefficients are
compared with experiments~\cite{Sydow98} at 19 MeV.  Again, we observe
appreciable effects introduced by the OPE-3NF diagram in the case of
the Bonn $B$ interaction, while for the CD-Bonn case the effects appear
to be smaller, although they go in the same direction.  We observe
that the calculation with Bonn $B$ + OPE-3NF
(Fig.~\ref{FIG-pd-bbonn-ndvtK-19}) is able to solve the
underprediction problem in $K_y^{x'x'-y'y'}$, but at the same time it
increases the discrepancies in $K_y^{z'z'}$, while in the case of the
CD-Bonn + OPE-3NF the situation remains essentially unchanged
(Fig.~\ref{FIG-pd-cdbonn-ndvtK-19}).

\section{Conclusions}
\label{Conclusions}

We have studied the low-energy effects in nucleon-deuteron
scattering due to an irreducible pionic effect leading to a
3$N$$F$ diagram of pronounced tensor structure. Overall we find
that the effects of this diagram on the differential cross section
are negligible in the considered energy range, indicating that the
average impact of this 3$N$$F$ diagram on 3$N$ dynamics is small.
On the other hand, for the case of the spin observables, the
higher sensitivity to the smaller components of the wave function
is able to detect the presence of this contribution.  This is so
especially for the case of the vector analyzing powers $A_y$ and
$iT_{11}$, which are considered to be a magnifying glass for the
triplet P waves of the 2$N$ subsystem. Independently of the 2$N$
interactions used as input, we found that this 3$N$$F$
contribution has the potential to significantly increase (about a
30\% effect) the magnitude of these two observables, and to solve
a discrepancy observed long time ago.  This is a consequence of
the specific spin-isospin structure of such 3$N$$F$ diagram which
affects in a privileged manner the triplet odd
states~\cite{Canton00,Canton01-b}.

For the remaining spin observables (tensor analyzing powers and
polarization transfer coefficients) the changes due to this pionic
contribution are small, however we found situations where the
diagram produces appreciable effects.  In particular with the use
of the Bonn B potential there are indications that the corrections
produced move towards the right direction, but the results with
the newer CD-Bonn potential reduce the size of these changes
considerably.  We found few cases where the calculations are
slightly - but appreciably - different when the 3$N$$F$ effect is
calculated with Bonn $B$ or CD-Bonn potentials. It happens with
$T_{21}$ in the energy range 10-20 MeV, with $K_y^{y'}$ in the
energy range 15-19 MeV, and with $K_y^{x'x'-y'y'}$ and
$K_y^{z'z'}$ around 19 MeV.  There are however still too many
uncertainties for arriving at a definite conclusion about these
slight differences.  Comparison of the wealth of experimental data
for charged particles has been done using only the Coulomb
slow-down hypothesis, while a consistent inclusion of Coulomb
effects is in principle required.  Finally, a more complete study
requires also the inclusion of the remaining 3$N$$F$ diagrams of
different topology. In particular, it is known that the 2$\pi$3NF
diagram produces additional corrections which are needed for
removing the underbinding of the 3$N$ bound state. The influence
on the vector analyzing powers in the considered energy range,
however, appears to be marginal.

\acknowledgements

This work is supported by the Italian MURST-PRIN Project ``Fisica
Teorica del Nucleo e dei Sistemi a Pi\`u Corpi".  W.~Sch. thanks
INFN and the University of Padova for hospitality and acknowledges
support from the Natural Science and Engineering Research Council
of Canada.

\appendix

\section{binding energy}

With the same model interactions used in the main text, we have
calculated also the triton binding energy. The results are given in
the Tab.~\ref{triton}.  For each one of the three potentials, the
first value in the table reports the 3$N$ binding energy calculated by
a direct solution of the homogeneous Faddeev equation in two momentum
variables (2D), using the original 2$N$ potential as input, and
without resorting to the separable expansion method. Details on the
computational method are explained in Ref.~\cite{Schadow98} and
references therein.  The second value has been calculated also via a
direct solution of the 2D Faddeev equation, but this time using as
input the separable expansion of the 2$N$ potential, with the same
ranks as given by Tab.~\ref{tab1}.  The third value represents the
binding energy obtained with the same separable representation of the
2$N$ potential as in the previous line, but using the one-dimensional
algorithm corresponding to the Lovelace-Alt-Grassberger-Sandhas
homogeneous equation~\cite{Schadow98}. Finally, the last line includes
the irreducible pionic effects, as discussed in this work, in the 1D
calculation for the binding energy.

\newpage

\begin{table}

\begin{center}

\begin{tabular}{|c|ccc|}
NN states & PEST  & BBEST & CDBEST \cr
\hline
$^1S_0 $       & 5 &     5 & 5(+5) \cr
$^3S_1-^3D_1 $ & 6 &     6 & 6 \cr
$^3P_0 $       & 5 &     4 & 4 \cr
$^1P_1 $       & 5 &     4 & 4 \cr
$^3P_1 $       & 5 &     4 & 4 \cr
$^1D_2 $       & 5 &     4 & 4 \cr
$^3D_2 $       & 5 &     4 & 4 \cr
$^3P_2-^3F_2 $ & 5 &     5 & 5 \cr
\end{tabular}

\end{center}

\caption[*]{Ranks used in the separable expansion for each 2$N$ state and
for the various potentials. For the CDBEST potential in the $^1S_0$
channel, the expansion has been done separately for $nn$ and $pn$ cases.}
\label{tab1}
\end{table}

\begin{table}

\begin{center}

\begin{tabular}{|c|cc|}
E (MeV)  & BBEST & CDBEST \cr \hline 3.0     &       0.730   &
0.58 \cr 8.5     &       0.733   &     0.60 \cr 11.3    &
0.743   &     0.63 \cr 13.3    &       0.753   &     0.65 \cr 15.3
&       0.763   &     0.67 \cr 17.3    &       0.773   &     0.69
\cr 18.3    &       0.778   &     0.70 \cr
\end{tabular}

\end{center}

\caption[*]{Energy dependence of the effective parameter
used to govern the cancellation in Eq.~(\ref{SUB-1}).}
\label{Parameters}

\end{table}

\begin{table}

\begin{center}

\begin{tabular}{|l|ccc|}
{\ } & Paris  & Bonn $B$ & CD Bonn \cr
\hline
 2D - orig       & -7.385   & -8.101   & -7.958     \cr
 2D - EST        & -7.376   & -8.088   & -7.947     \cr
 1D - EST        & -7.376   & -8.088   & -7.947     \cr
 1D + {\bf 3NF}  & -7.663   & -7.943   & -8.077     \cr
\end{tabular}

\end{center}

\caption[*]{Results obtained for the triton binding energy (MeV). The first
three lines correspond to different calculational methods without the
inclusion of the 3$N$$F$ effect. The last line includes the effects of
the OPE-3NF diagram. The 3$N$$F$ effect has been calculated with the effective
parameter $c$ determined at 3 MeV.}

\label{triton}

\end{table}

\newpage

\end{document}